\documentclass[11pt]{article}
\usepackage[english,activeacute]{babel}
\usepackage{natbib}
\usepackage{comment}
\usepackage{float}
\usepackage[hidelinks]{hyperref}
\usepackage{mathrsfs}
\usepackage{enumitem}
\usepackage[font={small,it}]{caption}
\usepackage{amsmath,amsfonts,amsthm,amssymb}
\usepackage{bm,rotating,multirow,dsfont,graphicx}
\usepackage[usenames, dvipsnames]{color}
\usepackage{url}
\usepackage{multicol}
\usepackage{multirow}
\usepackage[T1]{fontenc}
\usepackage{flafter}
\usepackage{appendix}
\usepackage{subfigure}
\usepackage{xcolor}
\usepackage{soul}
\makeatletter
\def\hlinewd#1{%
	\noalign{\ifnum0=`}\fi\hrule \@height #1 %
	\futurelet\reserved@a\@xhline}
\makeatother
\newcommand{\be}{\begin{eqnarray*}}
\newcommand{\ee}{\end{eqnarray*}}
\newcommand{\bet}{\begin{eqnarray}}
\newcommand{\eet}{\end{eqnarray}}
\def\spacingset#1{\renewcommand{\baselinestretch}{#1}\small\normalsize}\spacingset{1}
\def\@roman#1{\romannumeral #1}

\begin{document}

\title{Advances in Bayesian Modeling: Applications and Methods}

\date{}

\author{
    Yifei Yan, University of California, Santa Cruz, US \\
    Juan Sosa, Universidad Nacional de Colombia, Colombia \\
    Carlos A. Martínez, Universidad Nacional de Colombia, Colombia
}

\maketitle

\begin{abstract} 
This paper explores the versatility and depth of Bayesian modeling by presenting a comprehensive range of applications and methods, combining Markov chain Monte Carlo (MCMC) techniques and variational approximations. Covering topics such as hierarchical modeling, spatial modeling, higher-order Markov chains, and Bayesian nonparametrics, the study emphasizes practical implementations across diverse fields, including oceanography, climatology, epidemiology, astronomy, and financial analysis. The aim is to bridge theoretical underpinnings with real-world applications, illustrating the formulation of Bayesian models, elicitation of priors, computational strategies, and posterior and predictive analyses. By leveraging different computational methods, this paper provides insights into model fitting, goodness-of-fit evaluation, and predictive accuracy, addressing computational efficiency and methodological challenges across various datasets and domains.
\end{abstract}


\spacingset{1.1} 

\section{Introduction}

Bayesian inference has established itself as a cornerstone of modern statistical methodology, offering a framework for inference, prediction, and decision making. Its versatility lies in its ability to handle complex data structures, implicit uncertainty quantification (hence the colloquial quote ``uncertainty is for free''), and model comparison within a unified probabilistic framework. The rapid advancements in computational power and algorithms have enabled Bayesian methods to scale beyond their traditional limitations, fostering their application across a wide array of scientific disciplines, including oceanography, climatology, epidemiology, astronomy, and financial analysis, among many others.

At the heart of Bayesian computation lie two dominant approaches: Markov chain Monte Carlo (MCMC; e.g., \citealt{gamerman2006markov}) methods and variational inference (e.g., \citealt{blei2017}). MCMC methods have been extensively used due to their asymptotic properties when sampling from complex posterior distributions. Techniques like the Gibbs sampler and Metropolis-Hastings algorithm provide the foundation for a wide range of Bayesian analyses, including hierarchical and spatial models. However, their reliance on iterative sampling often results in high computational costs, making them less suitable for large-scale problems. In contrast, variational inference offers a deterministic alternative by framing posterior approximation as an optimization problem, trading some accuracy for significant gains in computational efficiency. This duality between MCMC and variational approaches has expanded the frontier of Bayesian methods, allowing researchers to design their computational strategies to the specific needs of their problems.

This paper explores Bayesian inference across a spectrum of applications and theoretical constructs, showcasing its adaptability to diverse data types and analytical challenges. The examples span hierarchical modeling for multi-level data structures, spatial modeling for geographic data, and higher-order Markov chains for complex sequential dependencies. Additionally, advanced topics such as Dirichlet process mixtures and Bayesian mixtures of Brownian motion are discussed to illustrate the ability of Bayesian methods to handle non-parametric and continuous-time processes. These examples highlight the dual role of Bayesian inference: As a theoretical tool for model development and as a practical solution for analyzing real-world data.

The applications in this work draw from datasets across various domains, each presenting unique challenges in data structure, computational requirements, and interpretability. For example, spatial modeling applied to climatology emphasizes the handling of geographic dependencies, while hierarchical models in oceanography address nested data structures. Bayesian methods are shown to provide robust solutions for model formulation, prior elicitation, computation, posterior analysis, predictive accuracy, and goodness-of-fit evaluation. Through this breadth of applications, the paper aims to bridge theoretical concepts with practical implementations, illustrating the versatility and depth of Bayesian methods.

The rest of the paper is structured as follows. Section 2 outlines the theoretical foundations of Bayesian inference and computational techniques, covering Markov chain Monte Carlo (MCMC) methods and variational approaches. Sections 3 to 6 explore specific applications, including hierarchical modeling, spatial modeling, higher-order Markov chains, and variational approximations for Dirichlet process mixtures. Each section details model formulation, computational strategies, and empirical results on domain-specific datasets. Finally, the discussion section synthesizes key findings and highlights directions for future research.

\section{A Brief Review on Bayesian Statistics}

This section introduces key concepts in Bayesian inference, focusing on the interpretation of probability and its role in statistical modeling. It explores the distinction between frequentist and Bayesian perspectives, highlighting how Bayesian methods incorporate prior beliefs and update them with observed data. Additionally, it discusses the computational challenges that arise in Bayesian analysis and the numerical techniques, such as Monte Carlo methods, that enable practical implementation in complex problems.

\subsection{Interpretations of Probability}

A probability measure is a sigma-additive measure defined on a sigma algebra of subsets of the sample space, taking values in the interval \([0,1]\). This axiomatic foundation has facilitated the development of a vast family of probability models applied across numerous fields. However, the interpretation of probability is crucial in the foundations of Bayesian inference. One perspective, known as the frequentist view, defines probability as the long-run relative frequency of an event, implying a conceptual framework where an experiment is repeated indefinitely under homogeneous conditions. In contrast, the subjective interpretation regards probability as a measure of personal belief or uncertainty. This perspective is particularly relevant in scenarios where repeated experimentation is infeasible, such as predicting the performance of a specific cyclist in an upcoming race.

The philosophical foundation of Bayesian inference has been shaped by seminal contributions from \cite{cox1946probability}, \cite{de1970logical}, and \cite{savage1972foundations}. A comprehensive exposition of the formal framework underpinning rational decision-making under uncertainty is provided by \cite[Chap. 2]{bernardo2009bayesian}. The key principle of rationality in this context lies in ensuring internal consistency within the decision-making process. Their work demonstrates that Bayes' theorem provides a principled mechanism for updating beliefs about the state of nature as new evidence becomes available.

\subsection{The trinity of Bayesian Inference: Notation and description of the problem}

To infer a (potentially multidimensional) parameter \( \pmb{\theta} \) from observed data \( \pmb{x} \), a Bayesian approach involves identifying plausible values for the components of \( \pmb{\theta} \) while quantifying the associated uncertainty. This uncertainty is represented through a probability distribution known as the prior distribution, which is specified \textit{before} any data is observed. The prior can incorporate existing knowledge and beliefs about the state of nature, in which case it is referred to as a ``subjective'' prior. Alternatively, it may encode minimal or no prior information about the parameter, in which case it is derived using general principles such as invariance under one-to-one transformations and is termed an ``objective,'' ``non-subjective,'' or ``default'' prior. In many cases, such priors are improper, meaning they do not integrate to one, but this does not necessarily hinder the validity of Bayesian inference. A thorough introduction to objective priors can be found in \cite{ghosh2011objective}.

The density or probability mass function of the prior distribution is denoted as \( p(\pmb{\theta}) \), with its associated parameters referred to as hyperparameters. In the Bayesian framework, all unknown model parameters are treated as random variables, making the specification of their probability distributions a fundamental step in model formulation. This process, known as prior elicitation, involves selecting a prior distribution that reflects existing knowledge, assumptions, or desired properties of \( \pmb{\theta} \) before observing any data.

Once data containing information about the parameter is observed, the focus shifts to updating knowledge and beliefs about it. This information is captured by the likelihood function, denoted as \( f(\pmb{x} \mid \pmb\theta) \), \( L(\pmb\theta) \), or \( L(\pmb{x}; \pmb\theta) \), depending on the notation used. The core principle of Bayesian inference establishes Bayes' rule as the optimal method for updating prior beliefs by combining the prior and the likelihood, resulting in the posterior distribution, which represents the updated distribution of the parameter given the observed data. The posterior density or mass function is denoted as \( p(\pmb\theta \mid \pmb{x}) \). Using this notation, Bayes' theorem is expressed as:  
\[
p(\pmb\theta \mid \pmb{x}) = \frac{p(\pmb{x} \mid \pmb\theta) \, p(\pmb\theta)}{p(\pmb{x})} \propto f(\pmb{x} \mid \pmb\theta) \, p(\pmb\theta),
\]
where \( p(\pmb{x}) \) is the marginal density or mass function of the data. According to \citet[Chap. 1]{bernardo2009bayesian}, although Bayes' theorem is a straightforward probability result that provides a principled way to learn about \( \pmb\theta \) from \( \pmb{x} \) while accounting for uncertainty, it has been a subject of controversy due to differing interpretations and the scope of its inputs. Nevertheless, the Bayesian approach has been shown to be both theoretically sound and highly applicable to real-world problems, as explored in this paper.

De Finetti’s representation theorem plays a fundamental role in Bayesian methods, as it establishes the existence of a prior distribution when the data are assumed to be exchangeable. This concept of exchangeability is particularly important because it naturally leads to key statistical modeling principles, such as independent random samples and parameter estimation \citep[Chap. 4]{bernardo2009bayesian}. The posterior distribution serves as the central object of inference in the Bayesian framework. It enables the estimation of parameters through point estimators, the construction of credible intervals that allow for probabilistic statements about \( \pmb\theta \), the testing of statistical hypotheses, the prediction of future observations, and the evaluation of model fit.

Large sample theory establishes that, under suitable regularity conditions, the posterior distribution exhibits asymptotic normality. Specifically, as the sample size increases, the posterior distribution can be approximated by a Normal distribution centered at the maximum likelihood estimator (MLE), with covariance given by the inverse of the observed Fisher information $\hat{\mathbf{I}}_n$. One of the most significant results in this context is the Bernstein-von Mises theorem, which is often regarded as a bridge between frequentist and Bayesian inference. This theorem implies that for large samples, the posterior mean closely approximates the MLE. More formally, under the regularity conditions of the Bernstein-von Mises theorem and assuming a prior with finite expectation, it can be shown that \citep{ghosh2007introduction}:  
\[
\sqrt{n} \left(\pmb{\theta}_n^\ast - \hat{\pmb{\theta}}_n^{\text{ML}} \right) \to 0, \quad \text{almost surely as } n \to \infty,
\]
where \( \pmb{\theta}_n^\ast \) and \( \hat{\pmb{\theta}}_n^{\text{ML}} \) the posterior mean and the MLE of \( \pmb{\theta} \), respectively.  

This result highlights the asymptotic equivalence of Bayesian and frequentist estimators. More precisely, in the Bayesian framework, we have:
\[
\hat{\mathbf{I}}_n^{1/2} (\pmb{\theta} - \hat{\pmb{\theta}}_n^{\text{ML}}) \mid \pmb{x} \to \textsf{N}_p(0, \mathbf{I}), \quad \text{almost surely},
\]
whereas in the frequentist setting:
\[
\hat{\mathbf{I}}_n^{1/2} (\pmb{\theta} - \hat{\pmb{\theta}}_n^{\text{ML}}) \mid \pmb{\theta} \to \textsf{N}_p(0, \mathbf{I}), \quad \text{almost surely}.
\]
These results illustrate that, under regularity conditions, Bayesian and frequentist inference converge asymptotically, reinforcing the robustness of Bayesian methods in large-sample scenarios.

\subsection{Toy Examples and the Need for Numerical Methods}

We present two toy examples to illustrate the computation of the posterior distribution and highlight the necessity of numerical methods in most real-life problems.  Consider the case where \( x_1, \dots, x_n \mid \theta \overset{\text{ind}}{\sim} \textsf{Bernoulli}(\theta) \) and the prior distribution is \( \theta \sim \textsf{Beta}(\alpha, \beta) \). This setup yields a product Bernoulli likelihood and a Beta prior with hyperparameters \( \alpha > 0 \) and \( \beta > 0 \). Applying Bayes' rule, the posterior distribution follows:  
\[
\theta \mid \pmb{x} \sim \textsf{Beta} \left( \alpha + \textstyle\sum_{i=1}^{n} x_i, \beta + n - \textstyle\sum_{i=1}^{n} x_i \right).
\]
A commonly used objective prior that corresponds to both Jeffreys’ prior (which is invariant under one-to-one transformations) and Bernardo’s reference prior (which maximizes the Kullback-Leibler divergence between the prior and the posterior) is \( \textsf{Beta}(1/2, 1/2) \). Another widely used prior is the Laplace prior, \( \textsf{Beta}(1,1) \), which corresponds to a uniform distribution on \( (0,1) \). The Haldane prior, defined by the improper distribution \( \alpha = \beta = 0 \), is also frequently considered. However, in this case, when all observed data points are either successes or failures, the resulting posterior distribution is improper. Therefore, when employing improper priors, it is essential to verify the properness of the posterior distribution to ensure valid inference. Now, suppose \( x_1, \dots, x_n \mid \theta \overset{\text{ind}}{\sim} \textsf{N}(\theta, \sigma^2) \) distribution, where \( \sigma^2 \) is known. Assuming a Normal prior \( \theta \sim \textsf{N}(\mu, \tau^2) \), the posterior distribution is given by  
\[
\theta \mid \pmb{x} \sim \textsf{N} \left( B \mu + (1 - B) \bar{x}, (1 - B)\tfrac{\sigma^2}{n} \right),
\]
where \( \bar{x} \) is the sample mean and $B = \sigma^2/(n\,\tau^2 + \sigma^2)$.

In both examples, the posterior distribution is available in closed form, allowing for direct computation of moments and other relevant quantities. These posterior characteristics can be used to derive point estimates, construct credible intervals, and assess the dispersion of the parameter of interest. From a decision-theoretic perspective, the choice of a point estimator—whether the posterior mean, median, or mode—is justified by minimizing the expected loss under different loss functions \citep{ghosh2007introduction}. For many real-life problems, posterior moments and even the posterior distribution itself cannot be derived in closed form. This limitation restricted the practical use of Bayesian inference for a long time. However, the advent of numerical algorithms and increased computing power enabled the approximation of posterior moments and other quantities of interest, such as the posterior probability of the parameter falling within a given set.

A major breakthrough in Bayesian computation came with the development of Monte Carlo methods, a family of techniques that approximate integrals and solve optimization problems by simulating random variables. These methods rely on strong convergence properties, ensuring accurate approximations as the number of simulated samples increases. While basic Monte Carlo methods are useful for relatively simple problems, more complex models—such as hierarchical structures—require more advanced approaches. In such cases, Markov chain Monte Carlo (MCMC) methods provide efficient sampling techniques to approximate posterior distributions.  The implementation of these methods will be demonstrated through various data analyses in this paper. However, a detailed exposition of Monte Carlo techniques is beyond the scope of this manuscript. For comprehensive treatments of the topic, the reader is referred to \cite{robert1999monte} and \cite{gamerman2006markov}.

\section{Bayesian Hierarchical Modeling of Sea Surface Temperature in the Mediterranean}

Sea surface temperature (SST) is an important indicator of environmental change, particularly in the context of global warming and marine feedback in the carbon cycle. The Mediterranean Sea, as one of the largest semi-closed European seas, serves as a valuable model for understanding broader oceanic changes. \cite{nykjaer2009mediterranean} indicated that the Mediterranean’s upper-layer temperature has been increasing by $0.03 \pm 0.008^\circ$C per year from 1985 to 2006, consistent with global warming trends. By analyzing SST data from different devices in December 2003, this section aims to provide insights into environmental changes in the Mediterranean and their implications for climate change \citep{joos1999global}.

Here, we present an analysis of SST measurements collected in the Mediterranean in December 2003 using four different device types across various locations. The dataset consists of 336 observations from 86 devices, including bucket, engine room intake (ERI), fixed buoys, and drifting buoys, with some devices providing replicated observations. The goal is to fit a Bayesian hierarchical model to account for the variability associated with device types, using MCMC to explore the posterior distribution. The analysis shows that the commonality of SST measurements varies depending on the device type, with the overall mean SST estimated to be 19.72$^\circ$C with a standard deviation of 0.65$^\circ$C.

\begin{table}[!b]
\centering
\begin{tabular}{l c c c c c c c}
\hline
 & \multicolumn{3}{c}{SST} & & \multicolumn{3}{c}{N\underline{o} of Records per device} \\
\cline{2-4} \cline{6-8}
 & N & Mean & SD & & Mean & Min & Max \\
\hline
Bucket & 36 & 19.3 & 1.5 & & 1.0 & 1 & 1 \\
D.Buoy & 10 & 19.9 & 0.9 & & 2.1 & 1 & 3 \\
ERI    & 35 & 19.6 & 1.4 & & 1.1 & 1 & 2 \\
F.Buoy & 5  & 20.4 & 0.9 & & 48.0 & 23 & 94 \\
\hline
Overall    & 86 & 19.6 & 1.4 & & 3.9 & 1 & 94 \\
\hline
\end{tabular}
\caption{Summary statistics of SST by device type.}
\label{tab_sst_1}
\end{table}

Among the 86 devices, 34 are bucket devices and 35 are ERIs, each representing approximately 40\% of the total. The remaining devices include 10 drifting buoys (d.buoys) and 5 fixed buoys (f.buoys). According to the descriptive statistics (Table \ref{tab_sst_1}) and the box plots (not shown here), the SSTs associated with different device types show slight variations. Fixed buoys have the highest mean and median SST among all groups. However, it is important to note that fixed buoys also have the fewest devices (only five) but the highest number of records per device, averaging 48 records.
The box plots indicate an outlier in the ERI group, originating from device 59. The SST reading of device 59 is 15.0$^\circ$C, the lowest among all devices.

\subsection{Modeling}

To assess the mean and variability of SST with device type as a predictor, we formulate a two-level Bayesian hierarchical model and use MCMC methods in order to estimate the model parameters.

\subsubsection{Bayesian hierarchical modeling}

Let \(y_{i,j}\) be the average SST measurement from the \(i\)-th device of the \(j\)-th device type, and \(n_{i,j}\) be the corresponding number of records associated with this device. We define a two-level hierarchical model in which the SST measurement \(y_{i,j}\) for each individual device is modeled as varying around the mean SST of its corresponding device type \(\theta_j\) (level 1), and the mean SSTs of the four device types vary around a global mean \(\mu\) (level 2).
Specifically, we assume that \(y_{i,j} = \theta_j + \epsilon_{i,j}\), where the $\epsilon_{i,j}$ are conditionally independent (as in the rest of the document), following \(\epsilon_{i,j}\mid\sigma^2_j \sim \textsf{t}_5(0, \sigma_j^2/n_{i,j})\), and \(\theta_j \mid \mu, \tau^2 \sim \textsf{N}(\mu, \tau^2)\). Here, \(j = 1 \ldots 4\) represents the device types: Bucket devices, drifting buoys, ERI, and fixed buoys, respectively, and the error variance within each device group is denoted by \(\sigma_j^2\).

On the one hand, note that at level 1, we model the errors using a Student’s \(\textsf{t}\) distribution with 5 degrees of freedom. This choice is motivated by the heavier tails of the \(\textsf{t}\) distribution compared to the Gaussian distribution, allowing for greater robustness to outliers and reducing their influence on the model estimates.
To facilitate sampling, we introduce auxiliary variables to represent such \(\textsf{t}_5\) distribution (see \citealt{albert1993bayesian} and \citealt[Chap. 12]{gelman2013bayesian}). Specifically, we introduce \(\lambda_{i,j}\) for each \(y_{i,j}\), such that \(y_{i,j} \mid \theta_j, \sigma_j^2, \lambda_{i,j} \sim \textsf{N}(\theta_j, \sigma_j^2 / (n_{i,j} \lambda_{i,j}))\), with \(\lambda_{i,j} \sim \textsf{G}(5/2, 5/2)\). Upon marginalization, it can be shown that this structure ensures that the errors follow a \(\textsf{t}_5\) distribution. On the other hand, we assume that all \(\sigma_j^2\) share a common scale \(\sigma^2\) and model them as \(\sigma_j^2 \mid \sigma^2 \sim \textsf{IG}(\alpha + 1, \alpha \sigma^2)\), with \(\sigma^2 \sim \textsf{G}(a, b)\). We set \(\alpha = 2\) to allow some variability in each \(\sigma_j^2\) relative to \(\sigma^2\), but not excessively. For instance, with \(\sigma^2 = 1\), about 90\% of the prior density for \(\sigma_j^2\) lies between 0.3 and 3. Additionally, we use \(a = b = 2\) to specify a relatively weak prior for \(\sigma^2\), centered at 1.

At level 2, we assume that the mean SSTs of each device type are drawn from a normal distribution centered around a global mean \(\mu\) with variance \(\tau^2\). Therefore, we model \(\theta_j \mid \mu, \tau^2 \sim \textsf{N}(\mu, \tau^2)\). Finally, we assign a flat prior for \((\mu, \tau^2)\), meaning that \(p(\mu, \log \tau^2) \propto 1\), or equivalently, \(p(\mu, \tau^2) \propto 1/\tau^2\) (\citealt[Chap. 3]{gelman2013bayesian}).

\subsubsection{Computation}

Let \( J = 4 \) be the number of device types, and let \( N_j \) be the number of devices of type \( j \). Then, the joint posterior is given by:

\begin{align*}
p(\pmb{\theta}, \pmb{\sigma}^2, \pmb{\lambda}, \mu, \tau^2, \sigma^2 \mid \pmb{y}) &\propto \prod_{j=1}^J \prod_{i=1}^{N_j} \frac{\lambda_{i,j}^{1/2}}{\sigma_j} \, \exp\left( -\frac{n_{i,j} \lambda_{i,j}}{2 \sigma_j^2} \, (y_{i,j} - \theta_j)^2 \right) \\
&\cdot \prod_{j=1}^J \prod_{i=1}^{N_j} \lambda_{i,j}^{5/2 - 1} \, \exp\left( -\frac{5}{2} \, \lambda_{i,j} \right) 
\cdot \prod_{j=1}^J \frac{1}{\tau} \, \exp\left( -\frac{1}{2 \tau^2} \, (\theta_j - \mu)^2 \right) \\
&\cdot \prod_{j=1}^J (\sigma_j^2)^{-(\alpha + 1)} \, \exp\left( -\frac{1}{\sigma_j^2} \, \alpha \, \sigma^2 \right) 
\cdot (\sigma^2)^{a - 1} \exp\left( -b \, \sigma^2 \right) \cdot \frac{1}{\tau^{2}}\,,
\end{align*}
where $\pmb{\theta} = (\theta_j)$, $\pmb{\sigma}^2 = (\sigma^2_j)$, $\pmb{\lambda} = (\lambda_{i,j})$, and $\pmb{y} = (y_{i,j})$.

We implement an MCMC algorithm to estimate the model parameters. The chain starts with 10,000 burn-in iterations, followed by 50,000 iterations with a thinning factor of 10, as determined from a test run. Thus, the final sample consists of 5,000 thinned simulations. The algorithm is based on the following update steps:
\begin{enumerate}
    \item Sample $\theta_j \mid \mu,\tau^{2}, \sigma_j^2,\lambda_{i,j} \sim \textsf{N}(\mu_{\theta_j},\sigma_{\theta_j}^{2})$, for $j=1,\ldots,J$, where
    $$
    \mu_{\theta_j} = 
    \frac{\frac{\mu}{\tau^{2}} + \sum_{j=1}^{N_j} \frac{n_{i,j}}
    {\sigma_j^{2}} \, \lambda_{i,j} \, y_{i,j}}{\frac{1}{\tau^{2}} + \sum_{j=1}^{N_j} \frac{n_{i,j}}
    {\sigma_j^{2}} \, \lambda_{i,j}}\,,\quad
    \sigma_{\theta_j}^{2} =
    \frac{1}{\frac{1}{\tau^{2}} + \sum_{j=1}^{N_j} \frac{n_{i,j}}
    {\sigma_j^{2}} \, \lambda_{i,j}}\,.
    $$

    \item Sample $\sigma_j^{2} \mid \theta_j,
        \lambda_{i,j},\sigma^{2}\sim \textsf{IG}(\alpha_{\sigma_j},\beta_{\sigma_j})$, for $j=1,\ldots,J$, where
    $$
    \alpha_{\sigma_j} = \alpha + \frac{N_j}{2}\,,\quad
    \beta_{\sigma_j} = \alpha\,\sigma^{2} + \frac{1}{2}\sum_{i=1}^{N_j} n_{ij}\,\lambda_{i,j}\, (y_{i,j}-\theta_j)^2\,.
    $$

    \item Sample $\lambda_{i,j} \mid \theta_j,\sigma_j^{2}\sim \textsf{G}(\alpha_{\lambda_{i,j}},
        \beta_{\lambda_{i,j}})$, for $i=1\ldots,N_j$ and $j=1\ldots J$, where
    $$
    \alpha_{\lambda_{i,j}} = 3\,,\quad 
    \beta_{\lambda_{i,j}} = \frac{5}{2} + \frac{1}{2}\, \frac{n_{i,j}}{\sigma_j^2}(y_{i,j} -\theta_j)^2\,.
    $$

    \item Sample  $\mu \mid \pmb\theta,\tau^{2}\sim\textsf{N}(\mu_\mu,\sigma^2_\mu)$, where
    $$
    \mu_\mu = \frac{1}{J}\sum_{j=1}^J\theta_j\,,\quad
    \sigma^2_\mu = \frac{\tau^2}{J}\,.
    $$

    \item Sample $\tau^{2}\mid\mu,\pmb\theta \sim \textsf{IG}(\alpha_{\tau},\beta_{\tau})$, where
    $$
    \alpha_{\tau} = \frac{J-1}{2}\,,\quad
    \beta_{\tau} = \frac{1}{2} \sum_{j=1}^J(\theta_j-\mu)^2\,.
    $$

    \item Sample $\sigma^{2} \mid \pmb\sigma^2 \sim \textsf{G}(\alpha_{\sigma}, \beta_{\sigma})$, where
    $$
    \alpha_{\sigma}=a\,,\quad
    \beta_{\sigma} = b + \alpha\sum_{j=1}^J \frac{1}{\sigma_j^{2}}\,.
    $$
\end{enumerate}

\subsection{Illustration}

In this case, the model quickly converges once the chain begins burning, and with a thinning factor of 20, the final sample comprises 5,000 simulations per model parameter. Across all variables in the final sample, the average effective sample size is 4,969, indicating negligible autocorrelation after thinning.

\begin{table}[!b]
\centering
\begin{tabular}{lcccccc}
\hline
Parameter & Mean & SD & 2.5\% & 50\% & 97.5\% \\
\hline
$\theta_1$ & 19.49 & 0.23 & 18.99 & 19.49 & 19.91 \\
$\theta_2$ & 19.75 & 0.25 & 19.28 & 19.73 & 20.29 \\
$\theta_3$ & 19.61 & 0.20 & 19.20 & 19.61 & 20.00 \\
$\theta_4$ & 20.04 & 0.38 & 19.39 & 20.01 & 20.80 \\
\hline
$\sigma_1$ &  1.27 & 0.18 &  0.96 &  1.25 &  1.67 \\
$\sigma_2$ &  1.04 & 0.27 &  0.64 &  0.99 &  1.69 \\
$\sigma_3$ &  1.27 & 0.18 &  0.89 &  1.16 &  1.55 \\
$\sigma_4$ &  3.95 & 1.52 &  1.75 &  3.70 &  7.50 \\
\hline
$\mu$      & 19.72 & 0.85 & 18.88 & 19.69 & 20.70 \\
$\tau$     &  0.65 & 1.16 &  0.02 &  0.42 &  2.67 \\
$\sigma$   &  0.50 & 0.19 &  0.18 &  0.49 &  0.89 \\
\hline
\end{tabular}
\caption{Posterior statistics of $\pmb\theta$, $\pmb\sigma$, $\mu$, $\tau$, and $\sigma$.}
\label{tab_sst_2}
\end{table}

\subsubsection{Parameter estimation}

Table \ref{tab_sst_2} summarizes the posterior distributions of the key parameters: \(\theta_1, \ldots, \theta_J\), \(\sigma_1, \ldots, \sigma_J\), \(\mu\), \(\tau\), and \(\sigma\). The posterior means for \(\theta_j\) range from 19.49\(^\circ\)C to 20.04\(^\circ\)C, reflecting a pattern consistent with Table \ref{tab_sst_1}. Specifically, the bucket device records the lowest mean, followed by ERI and drifting buoys, while floating buoys exhibit the highest mean. The posterior means for \(\mu\) and \(\tau\) are 19.72\(^\circ\)C and 0.65\(^\circ\)C, respectively. Moreover, histograms of the model parameters (not shown here) suggest that most parameters exhibit approximately symmetric, unimodal posterior distributions with relatively low variance. However, the variance components \(\tau^2\) and \(\sigma^2\) deviate from this pattern, displaying right-skewed posterior distributions with the majority of the probability mass concentrated near their respective posterior means.

\begin{figure}[!htb]\vspace{-4em}
    \centering
    \includegraphics[angle=-90,scale=0.9]{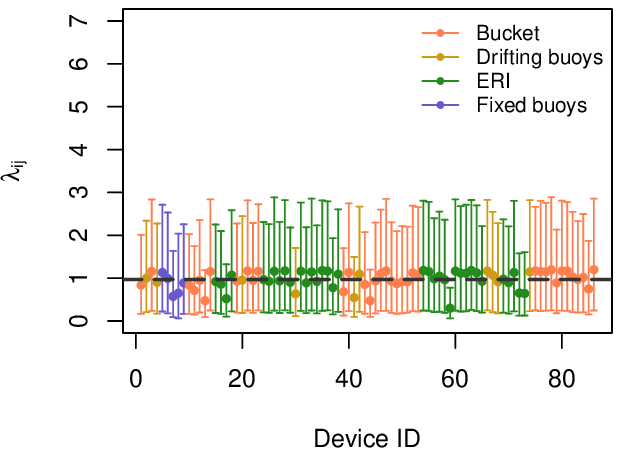}\vspace{-1em}
    \caption{Posterior means and 95\% quantile-based credible intervals for $\lambda_{i,j}$ across all devices, with the reference line indicating $\bar{\lambda}= 0.97$.}\label{fig_sst_1}
\end{figure}

Figure \ref{fig_sst_1} presents the posterior means and 95\% quantile-based credible intervals for $\lambda_{i,j}$ across all devices. The $\lambda_{i,j}$ values are generally on the same level, with reasonable 95\% credible intervals. The posterior means of $\lambda_{i,j}$ range from 0.30 to 1.20, with an average of 0.97 and a median of 0.98. No notable differences are observed between the $\lambda_{i,j}$ values across different device types.

\subsubsection{Commonality among device types}

Table \ref{tab_sst_2} highlights the heterogeneity in within-type variances. For the first three device types (bucket devices, drifting buoys, and ERI), the \(\sigma_j\) values are relatively close to 1. In contrast, \(\sigma_4\) for fixed buoys is notably higher, with a posterior mean of 3.95. This disparity is expected, as \(y_{i,j}\) represents the mean of all \(n_{i,j}\) records collected by device \((i,j)\), and its variance is given by \(\sigma_j^2 / n_{i,j}\). Consequently, larger \(\sigma_j\) values suggest greater variability in SST measurements within that device type, particularly for fixed buoys.

\begin{figure}[!htb]\vspace{-4em}
    \centering
    \includegraphics[angle=-90,scale=0.9]{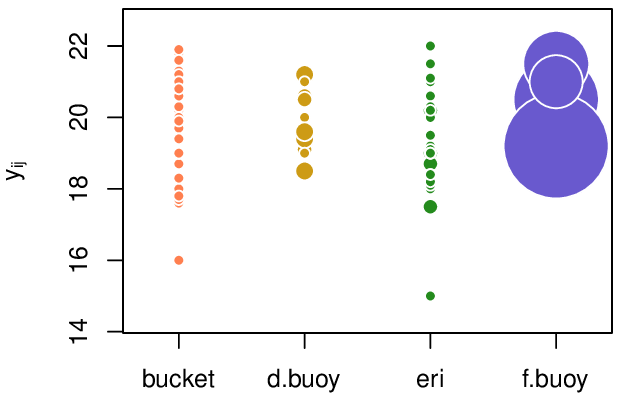}\vspace{-3em}
    \caption{$y_{i,j}$ (represented by the center of the circle) and the corresponding $n_{i,j}$ (proportional to the area of the circles) by device group.}\label{fig_sst_2}
\end{figure}

Figure \ref{fig_sst_2} displays the $y_{i,j}$ values (represented by the centers of the circles) and the corresponding $n_{i,j}$ values (indicated by the area of the circles) for each device group. In the fixed buoy group, the large circle areas indicate that each $y_{i,j}$ is derived from averaging a substantial number of records. As noted in Table \ref{tab_sst_1}, the average number of records per $y_{i,j}$ in the fixed buoy group is 48, nearly 50 times greater than that of the other three device types. Thus, while the $y_{i,j}$ values for fixed buoy devices seem closely clustered, this is due to averaging over large $n_{i,j}$ values. In fact, the variance within the fixed buoy group is not inherently small, as the $y_{i,j}$ values represent means calculated from a significant number of records.

A large variance within a device type indicates low commonality between individual devices of the same type. For device types such as bucket, drifting bucket, and ERI, where $\sigma_j$ values are approximately 1$^\circ$C, devices within each type exhibit some degree of shared measurement characteristics. In contrast, devices within the fixed buoy type show very weak commonality, as evidenced by $\sigma_4$ being around 4$^\circ$C.

\subsubsection{Outlier detection}

By employing the auxiliary variables $\lambda_{i,j}$ to reparameterize the $\textsf{t}_5$ distribution associated with the error terms $\epsilon_{i,j}$, we can utilize these $\lambda_{i,j}$ values for each device to identify potential outliers. Under this parameterization, the variance of $y_{i,j}$ is expressed as $\sigma_j^2/(n_{i,j}\lambda_{i,j})$. Consequently, a very small $\lambda_{i,j}$ corresponds to a large variance, indicating a significant deviation from the mean and thus signaling an outlier.

\begin{figure}[!htb]\vspace{-3em}
    \centering
    \includegraphics[angle=-90,scale=0.9]{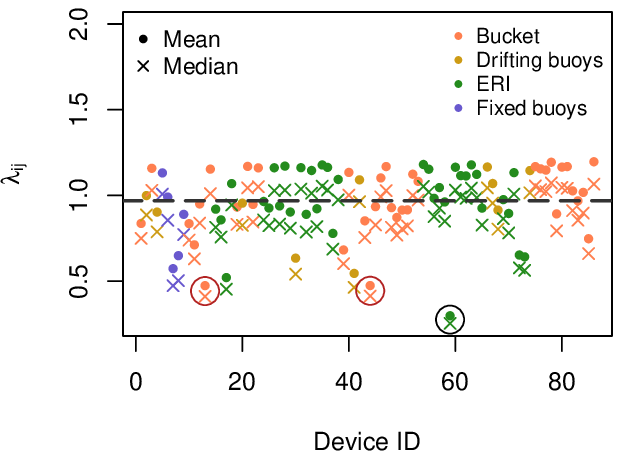}\vspace{-1em}
    \caption{Posterior mean and median of the $\lambda_{ij}$ values, with the reference line indicating $\bar{\lambda}= 0.97$.}\label{fig_sst_3}
\end{figure}

Figure \ref{fig_sst_3} shows the posterior mean and median of the $\lambda_{i,j}$ values. Only three devices have $\lambda_{i,j}$ values below 0.5: Device 59 (circled in black) and devices 13 and 44 (circled in red). Device 59 stands out with the lowest $\lambda_{i,j}$ value, having a mean of 0.30 and a median of 0.25, and is notably separated from all other $\lambda_{i,j}$ values, marking it as a clear outlier. In contrast, devices 13 and 44 also have very low $\lambda_{i,j}$ values but are not substantially distant from other low $\lambda_{i,j}$ values, classifying them as potential outliers rather than definitive ones.

\subsubsection{Goodness-of-fit}

We compute the Bayesian $p$-values (\citealt[Chap. 6]{gelman2013bayesian}) in order to evaluate the goodness-of-fit of the model. The Bayesian $p$-value evaluates the fit of a Bayesian model by comparing observed data \(\pmb{y}\) to the posterior predictive distribution of replicated data \(\pmb{z}\). Such value is defined as the probability that \(\pmb{z}\) is more extreme than \(\pmb{y}\) based on a test quantity \(T\), calculated as:
\[
    p_B = \Pr(T(\pmb{z}, \pmb{\theta}) \geq T(\pmb{y}, \pmb{\theta}) \mid \pmb{y}) = 
    \iint I(T(\pmb{z}, \pmb{\theta}) \geq T(\pmb{y}, \pmb{\theta})) \, p(\pmb{z} \mid \pmb{\theta}) \, p(\pmb{\theta} \mid \pmb{y}) \, \text{d}\pmb{z} \, \text{d}\pmb{\theta}\,,
\]
where \(I(\cdot)\) is the indicator function and $\pmb\theta$ are the model parameters. This probability incorporates uncertainty in both the posterior distribution of \(\pmb{\theta}\) and the posterior predictive distribution of \(\pmb{z}\), which provides a measure of model fit. Here, extreme values of the \(p\)-values are a sign of lack of model fit.

In this case, we predict the new observations of each device using the MCMC samples. In this way, for each \(y_{i,j}\) in the \(k\)-th MCMC sample, we generate \(z_{i,j}\) from
\[
    z_{i,j}^{(k)} \sim \textsf{N}(\theta_j^{(k)}, {\sigma_j^{2\,(k)}} / (n_{i,j} \lambda_{i,j}^{(k)}))\,,
\]
and we define the test quantity as \(T(z_{i,j}^{(k)}, \theta_j^{(k)}) = \lvert z_{i,j}^{(k)} - \theta_j^{(k)} \rvert\). Based on the predicted \(z_{i,j}\)'s for each device and each sample from the MCMC simulation, the average Bayesian \(p\)-value turns out to be 0.44. Therefore, we consider that our model fits the data relatively well.

\begin{figure}[!htb]\vspace{-3em}
    \centering
    \includegraphics[angle=-90,scale=0.9]{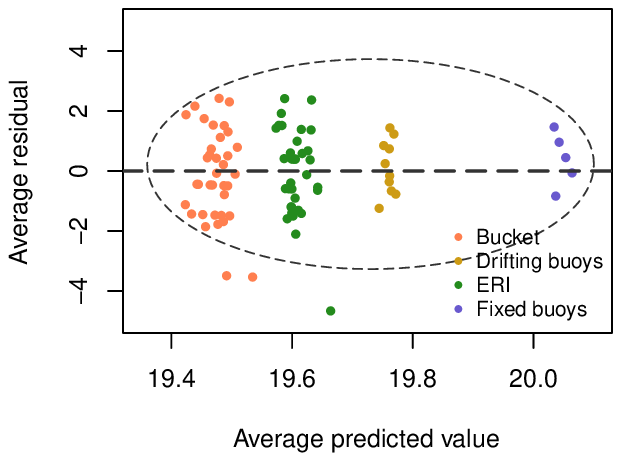}\vspace{-1em}
    \caption{Residual-predicted value plot.}\label{fig_sst_4}
\end{figure}

Alternatively, for a graphical check of the goodness-of-fit, we use a residual-predicted value plot, where the $x$-axis represents the average predicted value $z_{i,j}$ and the $y$-axis represents the average residual $y_{i,j} - z_{i,j}$. A healthy residual plot should have randomly scattered points with no specific patterns. Figure \ref{fig_sst_4} shows the residual-predicted value plot, in which we observe that the residuals exhibit a healthy elliptical distribution with no discernible pattern. The three points detached from the others correspond to the outlier (device 59, displayed in green) and the potential outliers (devices 13 and 44, displayed in coral) that we identified previously.

\subsubsection{Remarks}

Under our Bayesian hierarchical model, the mean SST of certain areas of the Mediterranean in December 2013 is estimated to be \(19.72 \pm 0.65\)\,\(^\circ\text{C}\). The variability among devices of the same type differs depending on the device type. More detailed data would enable more accurate estimation of both the between- and within-device-type variation in SST measurements.

\section{Bayesian Spatial Modeling of Rainfall Data}

The temporal and geographic distribution of precipitation provides valuable insights into the water balance and the impacts of climate change \citep{sarmiento2000water}. Understanding rainfall patterns and making location- and time-specific precipitation predictions are critical challenges in agriculture and ecology. In this section, we present a Bayesian hierarchical spatial model to analyze rainfall data collected between 1968 and 1983 in Guárico, one of the 23 states in north-central Venezuela. The dataset also includes the longitude, latitude, altitude of each collection site. Bordered to the north by the central highlands and to the south by the Orinoco River, Guárico spans 25,091 square miles (64,986 square kilometers) of predominantly plains, making it the leading rice-producing state in Venezuela. Analyzing Guárico's rainfall offers valuable insights for local agricultural practices and aids in developing scalable models to better understand the patterns and predictors of annual precipitation.

\begin{figure}[!htb]
    \centering
    \includegraphics[scale=0.75]{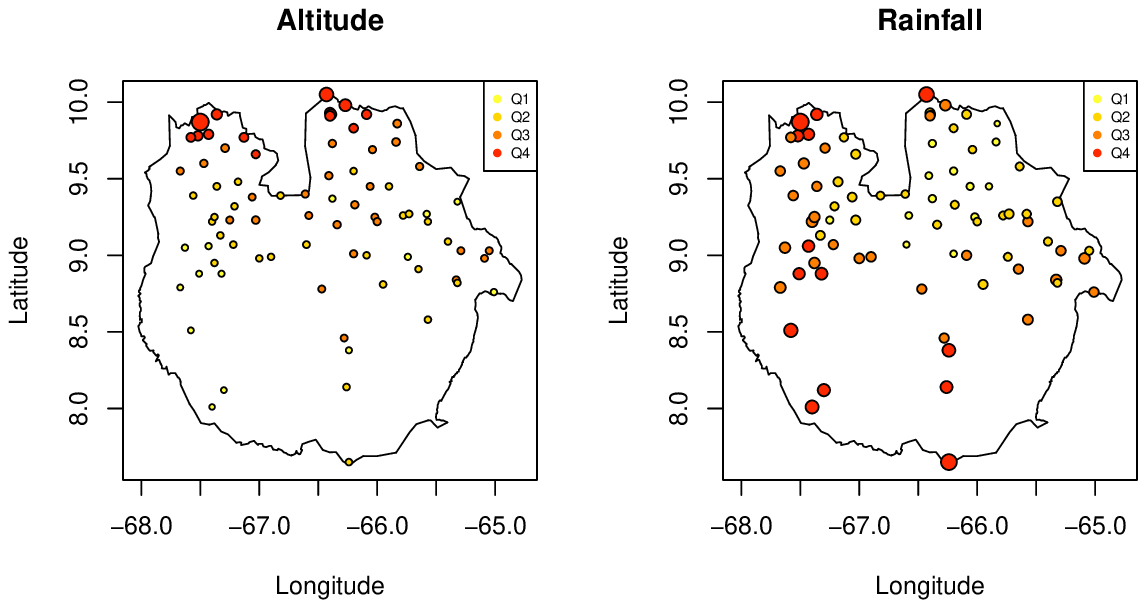}
    \caption{Latitude and mean rainfall across all years by location.}
    \label{fig_sp_1a}
\end{figure}

Table \ref{tab_sp_1} provides a summary of rainfall data for all years and grouped into four-year intervals, revealing noticeable temporal variation over time. This variation suggests the need to account for temporal factors when constructing the model. Furthermore, Figure \ref{fig_sp_1a} illustrates the latitude and mean rainfall across all years by location, showing an increase in elevation along the northern edge of Guárico. The data indicate heavier precipitation in the central southern region and along the northern border compared to other areas of the state.

\begin{table}[!htb]
\centering
\begin{tabular}{lccccccc}
\hline
 & \multicolumn{2}{c}{Records} & & \multicolumn{4}{c}{Rainfall} \\
 \cline{2-3} \cline{5-8}
 & N & Missing & & Mean & SD & 2.5\% & 97.5\% \\
\hline
All years      & 1280  & 190 &  & 1047 & 330  & 558  & 1913  \\
1968 - 1971    & 320   & 74  &  & 1200 & 331  & 742  & 1927  \\
1972 - 1975    & 320   & 34  &  & 850  & 219  & 468  & 1323  \\
1976 - 1979    & 320   & 27  &  & 1042 & 276  & 645  & 1688  \\
1980 - 1983    & 320   & 55  &  & 1066 & 318  & 259  & 1688  \\
\hline
\end{tabular}
\caption{Summary statistics of rainfall by years.}
\label{tab_sp_1}
\end{table}

\begin{figure}[!htb]
    \centering
    \includegraphics[scale=0.75]{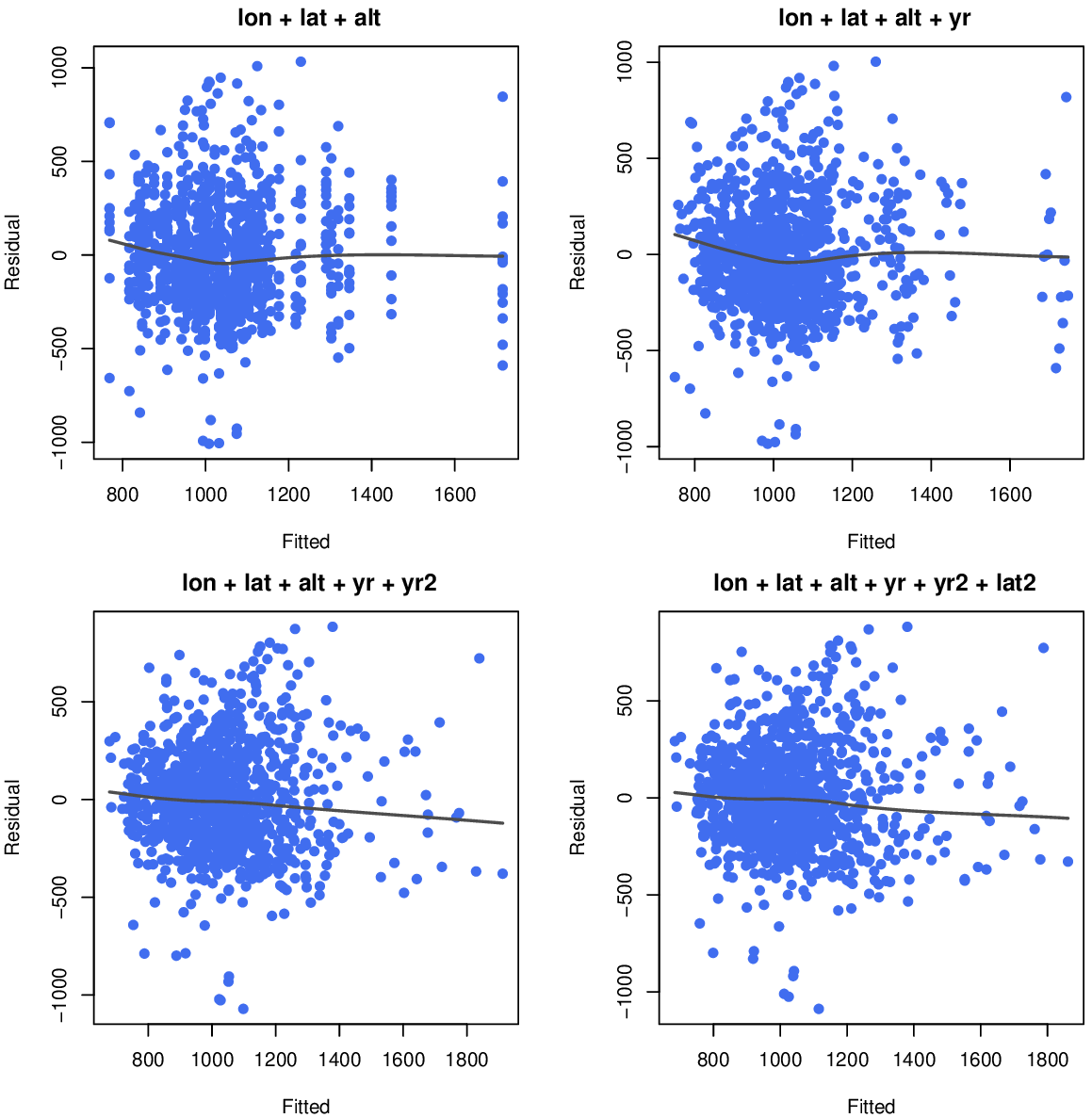}
    \caption{Residuals versus predicted values of linear models using mean rainfall as the response variable and several combination of predictors as explanatory variables.}
    \label{fig_sp_1b}
\end{figure}

To identify a suitable trend function for the model, we fitted several linear models with mean rainfall as the response variable and several combinations of predictors as explanatory variables. Figure \ref{fig_sp_1b} presents the residuals versus predicted value plots used to compare these models and evaluate their fit. Incorporating first-order terms for all variables, along with quadratic terms for year and latitude, substantially improved the model's fit and resulted in well-distributed residuals. In contrast, other second-order terms offered minimal explanatory value and occasionally introduced residual clustering. Based on this analysis, we selected a trend function that includes all first-order terms and quadratic terms for year and latitude to effectively capture the spatial and temporal dynamics.

\subsection{Modeling}

To analyze the influence of predictors on precipitation and predict rainfall for a given year, we develop a Bayesian hierarchical spatial model. Parameter estimation is performed using an MCMC approach with a block sampler, ensuring efficient computation and reliable inference.

\subsubsection{Bayesian hierarchical spatial modeling}

\begin{figure}[!htb]
    \centering
    \includegraphics[scale=0.75]{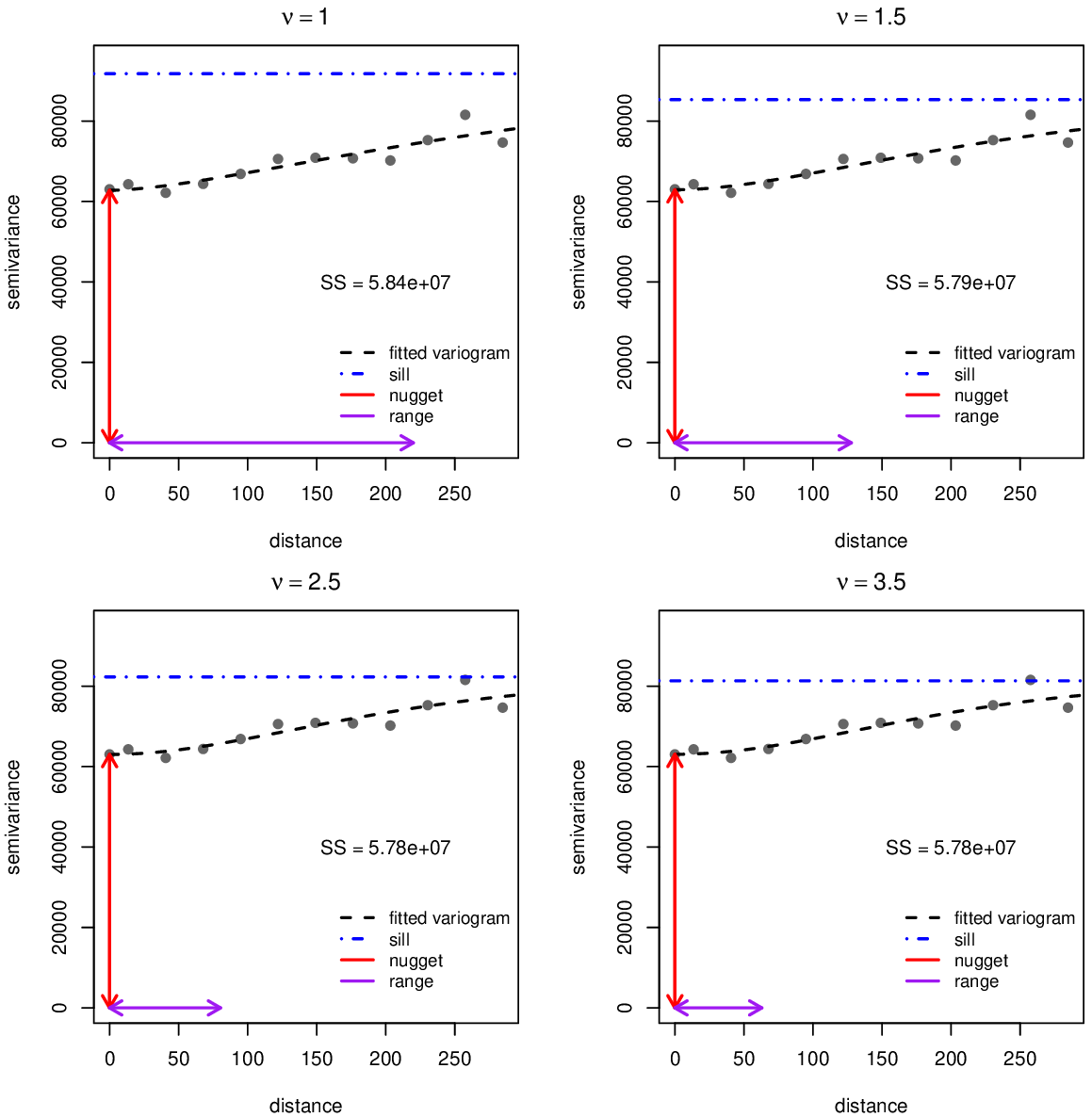}
    \caption{Least square fit of variagram using different values of $\nu$.}
    \label{fig_sp_1c}
\end{figure}

Let \(\mathbf{X} = (\mathbf{X}_o, \mathbf{X}_\ast)\) represent the complete dataset, comprising both observed and missing rainfall values, and let \(\mathbf{D} = (\mathbf{D}_o, \mathbf{D}_\ast)\) denote the corresponding trend matrix, which includes the predictors: \texttt{longitude}, \texttt{latitude}, \texttt{altitude}, \texttt{year}, \texttt{latitude$^2$}, and \texttt{year$^2$}. We assume that \(\mathbf{X}\) follows a Gaussian process characterized by the following mean and covariance structure:  
\[
    \mathbf{X} \sim \textsf{N}(\mathbf{D} \pmb{\beta}, \sigma^2 \mathbf{R}(\phi) + \tau^2 \mathbf{I}),
\]
where \(\pmb{\beta}\) represents the vector of coefficients of the predictors, \(\sigma^2\) is the variance of the underlying process, \(\tau^2\) is the nugget effect (including observational
error), \(\mathbf{R}(\phi)\) is a Matérn correlation matrix defined by the range parameter \(\phi\) and a specified order \(\nu\), and \(\mathbf{I}\) denotes the identity matrix. For a detailed explanation of these spatial modeling fundamentals, refer to \cite{Banerjee2015}. For this analysis, \(\nu\) is fixed at 2.5, selected based on a comparison of the least-squares fit of variograms under Matérn covariance models with varying \(\nu\) values (Figure \ref{fig_sp_1c}). A graphical inspection and the corresponding least-squares values indicate that increasing \(\nu\) beyond 2.5 provides negligible improvement to the fit.

For convenience in inference and sampling, we make \(\gamma^2 = \tau^2 / \sigma^2\) and rewrite the variance as \(\sigma^2 \mathbf{V}(\phi, \gamma^2)\), where \(\mathbf{V}(\phi, \gamma^2) = \mathbf{R}(\phi) + \gamma^2 \mathbf{I}\). Using this reparameterization, we formulate the following Bayesian hierarchical spatial model:
\[
\begin{aligned}
    \mathbf{X} \mid \pmb{\beta}, \sigma^2, \gamma^2, \phi &\sim \textsf{N}(\mathbf{D} \pmb{\beta}, \sigma^2 \mathbf{R}(\phi) + \tau^2 \mathbf{I}) \\
    p(\pmb{\beta}, \sigma^2) &\sim 1/\sigma^2 \\
    p(\phi, \gamma^2) &\sim 1/\phi \cdot \textsf{U}(\gamma^2 \mid 2, 4)\,.
\end{aligned}
\]

The variogram suggests that the dataset exhibits a significant nugget effect, corresponding to a relatively large \(\gamma^2\). As a result, we assign a uniform prior to \(\gamma^2\) over \((2, 4)\). Noninformative reciprocal priors are chosen for \(\sigma^2\) and \(\phi\).

\subsubsection{Computation}

Let $n$ and $p$ be the number of data points and the number of predictors, respectively. Under the model given above, the posterior distribution is given by:
\[
\begin{aligned}
    p(\mathbf{X}, \pmb{\beta}, \sigma^2, \gamma^2, \phi) \propto & \, (\sigma^2)^{-(n-p)/2}\, |\mathbf{V}(\phi, \gamma^2)|^{-1/2} \\
    & \cdot \exp\bigg\{-\frac{1}{2\sigma^2} \Big[(\pmb{\beta} - \hat{\pmb{\beta}})^\top \mathbf{D}^\top \mathbf{V}^{-1}(\phi, \gamma^2) \mathbf{D} (\pmb{\beta} - \hat{\pmb{\beta}}) + S^2(\phi, \gamma^2)\Big]\bigg\} \\
    & \cdot (\sigma^2)^{-1} \, \phi^{-1} \, I(2 < \gamma^2 < 4)\,,
\end{aligned}
\]
where \(\hat{\pmb{\beta}}\) is a solution to 
$$
\mathbf{D}^\top \mathbf{V}^{-1}(\phi, \gamma^2) \mathbf{D} \hat{\pmb{\beta}} = \mathbf{D}^\top \mathbf{V}^{-1}(\phi, \gamma^2) \mathbf{X}\,,
$$
$S^2(\phi, \gamma^2) = (\mathbf{X} - \mathbf{D} \hat{\pmb{\beta}})^\top \mathbf{V}^{-1}(\phi, \gamma^2) (\mathbf{X} - \mathbf{D} \hat{\pmb{\beta}})$, and \(I(\cdot)\) is the indicator function. Recall that \(\hat{\pmb{\beta}}\) is unique if and only if $\mathbf{D}$ has full column rank.

Note that \(\mathbf{X}\) also includes \(\mathbf{X}_\ast\), which represents the latent variables corresponding to the missing values. The algorithm consists of a block sampler for all model parameters \((\pmb{\beta}, \sigma^2, \gamma^2, \phi)\) and a Gibbs step for \(\mathbf{X}_\ast\) conditioned on all other parameters.

\textit{Block sampler for $(\pmb\beta,\sigma^2,\gamma^2,\phi)$ given $\mathbf{X}_\ast$}

Define \(\pmb{\psi} = (\phi, \gamma^2)\). If we consider the following prior \(p(\pmb{\beta}, \sigma^2, \pmb{\psi}) \propto p(\pmb{\psi}) \, \textsf{IG}(\sigma^2 \mid a, b)\), it can be shown that the posterior distribution can be rewritten in a product form as follows:
\[
\begin{aligned}
    p(\pmb{\beta}, \sigma^2, \pmb{\psi} \mid \mathbf{X}) &=
    p(\pmb{\beta} \mid \sigma^2, \pmb{\psi}, \mathbf{X}) \,
    p(\sigma^2 \mid \pmb{\psi}, \mathbf{X}) \,
    p(\pmb{\psi} \mid \mathbf{X})\,,
\end{aligned}
\]
where
\[
\begin{aligned}
    p(\pmb{\beta} \mid \sigma^2, \pmb{\psi}, \mathbf{X}) &\propto
    \exp\left\{
    -\frac{1}{2\sigma^2} (\pmb{\beta} - \hat{\pmb{\beta}})^\top \mathbf{D}^\top \mathbf{V}^{-1}(\pmb{\psi}) \mathbf{D} (\pmb{\beta} - \hat{\pmb{\beta}})
    \right\} \\
    p(\sigma^2 \mid \pmb{\psi}, \mathbf{X}) &\propto
    (\sigma^2)^{-\left(\frac{n-k}{2} + a + 1\right)}
    \exp\left\{-\frac{1}{2\sigma^2} (S^2(\pmb{\psi}) + b)\right\} \\
    p(\pmb{\psi} \mid \mathbf{X}) &\propto
    |\mathbf{V}(\pmb{\psi})|^{-\frac{1}{2}}
    |\mathbf{D}^\top \mathbf{V}(\pmb{\psi}) \mathbf{D}|^{-\frac{1}{2}}
    (S^2(\pmb{\psi}) + 2b)^{-\left(\frac{n-k}{2} + a\right)} \, p(\pmb{\psi})\,,
\end{aligned}
\]
with \(S^2(\pmb{\psi}) = (\mathbf{X} - \mathbf{D} \hat{\pmb{\beta}})^\top \mathbf{V}^{-1}(\pmb{\psi}) (\mathbf{X} - \mathbf{D} \hat{\pmb{\beta}})\). The above also applies to a reciprocal prior for \(\sigma^2\) by setting \(a = 0\) and \(b = 0\), such that \(p(\sigma^2) \propto 1/\sigma^2\). This factorization suggests that we can apply a blocking strategy to sample from the posterior.

We choose a jumping distribution \(g_1\) for \(\pmb{\psi}\) and define the block jumping distribution as:
\[
g(\pmb{\beta}, \sigma^2, \pmb{\psi}) = p(\pmb{\beta} \mid \sigma^2, \pmb{\psi}, \mathbf{X}) \,
p(\sigma^2 \mid \pmb{\psi}, \mathbf{X}) g_1(\pmb{\psi})\,,
\]
where the acceptance ratio of the current sample \(\pmb{\psi}^{(i)}\) and the proposed sample \(\pmb{\psi}^\ast\) is:
\[
\begin{aligned}
    r = \frac{p(\pmb{\beta}^\ast, {\sigma^2}^\ast, \pmb{\psi}^\ast \mid \mathbf{X})}{
        p(\pmb{\beta}^{(i)}, {\sigma^2}^{(i)}, \pmb{\psi}^{(i)} \mid \mathbf{X})}
        \cdot \frac{g(\pmb{\beta}^{(i)}, {\sigma^2}^{(i)}, \pmb{\psi}^{(i)})}{
        g(\pmb{\beta}^\ast, {\sigma^2}^\ast, \pmb{\psi}^\ast)}
    = \frac{p(\pmb{\psi}^\ast \mid \mathbf{X}) \, g_1(\pmb{\psi}^{(i)})}{
        p(\pmb{\psi}^{(i)} \mid \mathbf{X}) \, g_1(\pmb{\psi}^\ast)}\,.
\end{aligned}
\]

Based on the above result, we only sample \((\pmb{\beta}, \sigma^2)\) if the proposed \(\pmb{\psi}^\ast\) is accepted. Conditioning on \(\pmb{\psi}\), we first sample \(\sigma^2\), then sample \(\pmb{\beta}\) conditioning on both \(\sigma^2\) and \(\pmb{\psi}\). Thus, we sample \((\pmb{\beta}, \sigma^2)\) as follows:
\begin{enumerate}
    \item Sample \(\sigma^2 \mid \pmb{\psi} \sim \textsf{IG}((n-k)/2, S^2(\pmb{\psi})/2)\).

    \item Sample \(\pmb{\beta} \mid \sigma^2, \pmb{\psi} \sim \textsf{N}(\hat{\pmb{\beta}}, \sigma^2 \mathbf{V}_\beta)\), where
    \[
    \hat{\pmb{\beta}} = \mathbf{V}_\beta \mathbf{D}^\top \mathbf{V}^{-1}(\pmb\psi) \mathbf{X}\,,
    \quad
    \mathbf{V}_\beta = \left(\mathbf{D}^\top \mathbf{V}^{-1}(\pmb{\psi}) \mathbf{D}\right)^{-1}\,.
    \]
\end{enumerate}

In practice, instead of inverting matrices, we use Cholesky decomposition and QR decomposition (e.g., \citealt{golub2013matrix}) to obtain \(\hat{\pmb{\beta}}\) and sample from the posterior of \(\pmb{\beta}\). Moreover, since the sample size is large, to facilitate sampling, we sample \(\pmb{\psi}\) using a grid approach (e.g., \citealt[Chap. 2]{gelman2013bayesian}), which is analogous to applying independent discrete priors. By doing this, we gain efficiency and stability in computation at the cost of a coarse resolution when exploring the posterior density of \(\pmb{\psi}\).

\textit{Gibbs step of missing values $\mathbf{X}_\ast$ given $(\pmb\beta,\sigma^2,\gamma^2,\phi)$}

It follows that \((\mathbf{X}_\ast, \mathbf{X}_o)\) has a joint Normal distribution. In this way, we rearrange the variance matrix into the following form:
\[
\tilde{\mathbf{V}} =
\begin{pmatrix}
    \tilde{\mathbf{V}}_{\ast\ast} & \tilde{\mathbf{V}}_{\ast o} \\[6pt]
    \tilde{\mathbf{V}}_{o\ast} & \tilde{\mathbf{V}}_{oo} \\
\end{pmatrix}\,,
\]
where \(\tilde{\mathbf{V}}_{\ast\ast}\) and \(\tilde{\mathbf{V}}_{oo}\) are the correlation matrices of \(\mathbf{X}_\ast\) and \(\mathbf{X}_o\), respectively, and \(\tilde{\mathbf{V}}_{\ast o}\) is the correlation matrix between \(\mathbf{X}_\ast\) and \(\mathbf{X}_o\). To update the latent variables and impute missing values, we sample \(\mathbf{X}_\ast \mid \mathbf{X}_o, \pmb{\beta}, \sigma^2, \pmb{\psi} \sim \textsf{N}(\pmb{\mu}_\ast, \pmb{\Sigma}_\ast)\), where:
\[
\pmb{\mu}_\ast = \mathbf{D}_\ast \pmb{\beta} + \tilde{\mathbf{V}}_{\ast o} \tilde{\mathbf{V}}_{oo}^{-1}
        (\mathbf{X}_o - \mathbf{D}_o \pmb{\beta})\,,\quad
\pmb{\Sigma}_\ast = \sigma^2\left(\tilde{\mathbf{V}}_{\ast\ast} - \tilde{\mathbf{V}}_{\ast o}
        \tilde{\mathbf{V}}_{oo}^{-1} \tilde{\mathbf{V}}_{o\ast}\right),
\]
where \(\mathbf{D}_o\) and \(\mathbf{D}_\ast\) are the corresponding design matrices of \(\mathbf{X}_o\) and \(\mathbf{X}_\ast\), respectively.

\subsection{Illustration}

We implement a MCMC algorithm to estimate the model parameters. The chain begins with 500 burn-in iterations to allow it to converge to the target distribution, followed by 1,000 effective iterations used for parameter estimation. No convergence issues are detected after running the chain, and therefore we are confident that the algorithm has successfully converged to the corresponding target distribution.

\subsubsection{Parameter Estimation}

Table \ref{tab_sp_2} presents the posterior distribution of the model parameters. The variables \texttt{longitude}, \texttt{latitude}, and \texttt{altitude} are scaled in kilometers, while the years are recoded from 1 to 16. The results reveal a strong positive trend in precipitation within the trend function as altitude increases. Specifically, for every 100-meter rise in altitude, the annual rainfall increases by an average of 86.5 millimeters. This observation is consistent with the heavy rainfall patterns observed in northern Gu{\'a}rico, located at the edge of the central highlands.

\begin{table}[!htb]
\centering
\begin{tabular}{lccccc}
\hline
Parameter              & Mean    & SD     & 2.5\%  & 50\%   & 97.5\% \\ \hline
\texttt{longitude}     & -0.56   & 0.054  & -1.84  & 0.55   & 0.39   \\ 
\texttt{latitude}      & -16.7   & 0.25   & -36.2  & -16.7  & 2.3    \\ 
\texttt{altitude}      & 865     & 81.8   & 748    & 859    & 1037   \\ 
\texttt{year}          & -86.3   & 5.96   & -97.5  & -86.2  & -73.1  \\ 
\texttt{latitude}$^2$  & 0.01    & 0.01   & 0.00   & 0.01   & 0.02   \\ 
\texttt{year}$^2$      & 4.78    & 0.34   & 4.05   & 4.78   & 5.44   \\
$\sigma^2$             & 17098   & 1005   & 15444  & 17027  & 19245  \\ 
$\phi$                 & 49.8    & 8.63   & 35.0   & 47.0   & 65.0   \\ 
$\gamma^2$             & 3.25    & 0.16   & 3.00   & 3.25   & 3.50   \\ \hline
\end{tabular}
\caption{Posterior statistics of the model parameters.}\label{tab_sp_2}
\end{table}

The posterior estimates also indicate a second-order trend in latitude and year, aligning with the observation that higher precipitation levels are concentrated in both the southern and far northern regions of Gu{\'a}rico. Geographically, southern Gu{\'a}rico is bordered by the Orinoco River, a natural source of rainfall. Moreover, the central southern region encompasses the Aguaro-Guariquito National Park, where extensive forest coverage promotes water conservation and creates conditions conducive to abundant rainfall.

The posterior mean of \(\gamma^2\) is 3.25, reflecting a significant nugget effect and indicating substantial observational error, likely influenced in part by inaccuracies in the manual recording of rainfall data. Furthermore, the formulation of the trend function impacts the distribution of variability between the trend and the spatial component, which may contribute to the estimation of \(\gamma^2\). The posterior mean of the range parameter \(\phi\) is 49.8, implying a relatively gradual decay in spatial correlation.

\subsubsection{Goodness-of-fit}

To assess model fit, we perform validation by excluding two randomly selected observations from the dataset. The model is then refitted using the remaining data. Next, we generate samples from the posterior predictive distribution for the excluded observations and compare these predictions with the observed values. If the true values lie close to the center of the predictive distribution, it provides evidence that the model captures the underlying data structure effectively and fits the dataset well.

\begin{figure}[!htb]
    \centering
    \includegraphics[scale=0.75]{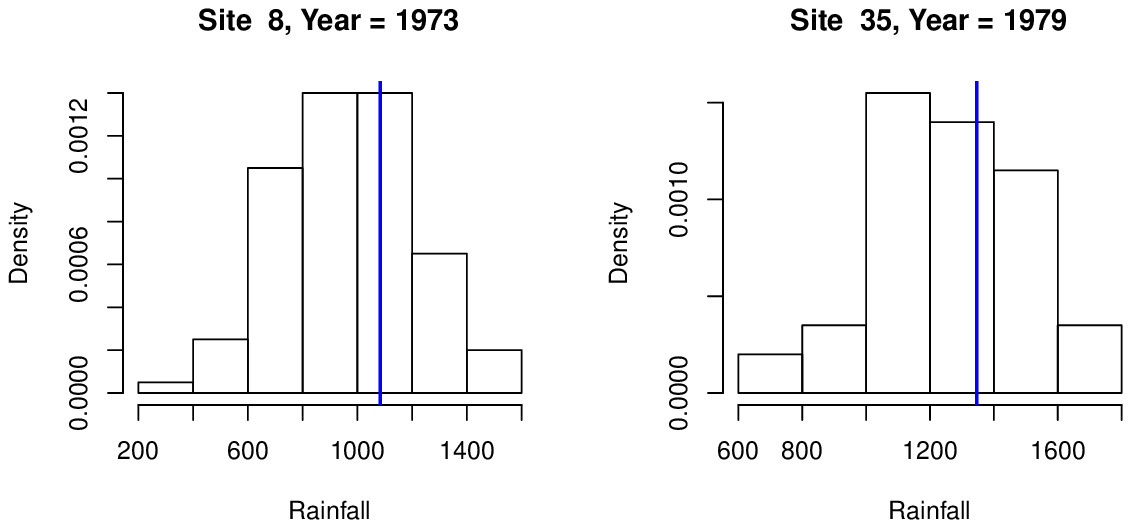}
    \caption{Model checking.}
    \label{fig_sp_2b}
\end{figure}

Figure \ref{fig_sp_2b} presents the histograms of the posterior predicted rainfall for the two hold-out observations, with the true values indicated by blue vertical lines. Both posterior predictive distributions exhibit an approximately symmetric and unimodal shape, and in both cases, the true values are positioned near the centers of their respective distributions. This alignment suggests that the model provides a good fit to the data.

\subsubsection{Predictions}

Predictions at new locations with a covariate matrix \(\mathbf{\hat{D}}\) are jointly Normal with the complete data \(\mathbf{X}\). For each sample of \(\mathbf{X}\) (including a sample of \(\mathbf{X}_\ast\)) and the parameter vectors, we draw \(\mathbf{\hat{X}} \sim \textsf{N}(\pmb{\mu}_{\mathrm{new}}, \pmb{\Sigma}_{\mathrm{new}})\), where
\[
\pmb{\mu}_{\mathrm{new}} = \mathbf{\hat{D}}\pmb{\beta} + \mathbf{R}\mathbf{V}^{-1}(\mathbf{X} - \mathbf{D}\pmb{\beta}), \quad 
\pmb{\Sigma}_{\mathrm{new}} = \sigma^2(\mathbf{\hat{V}} - \mathbf{R}\mathbf{V}^{-1}\mathbf{R})\,,
\]
with \(\mathbf{\hat{V}}\) representing the correlation matrix of \(\mathbf{\hat{X}}\), and \(\mathbf{R}\) denoting the correlation matrix between \((\mathbf{\hat{X}}, \mathbf{X})\). To further analyze predictions, we compute the probability of rainfall exceeding 1200 millimeters, which corresponds to the third quartile of all non-missing observations, using the predictive samples. Since altitude data at the new location is unavailable, we impute the altitudes using those from the nearest collection site.

\begin{figure}[!t]
    \centering
    \includegraphics[scale=0.72]{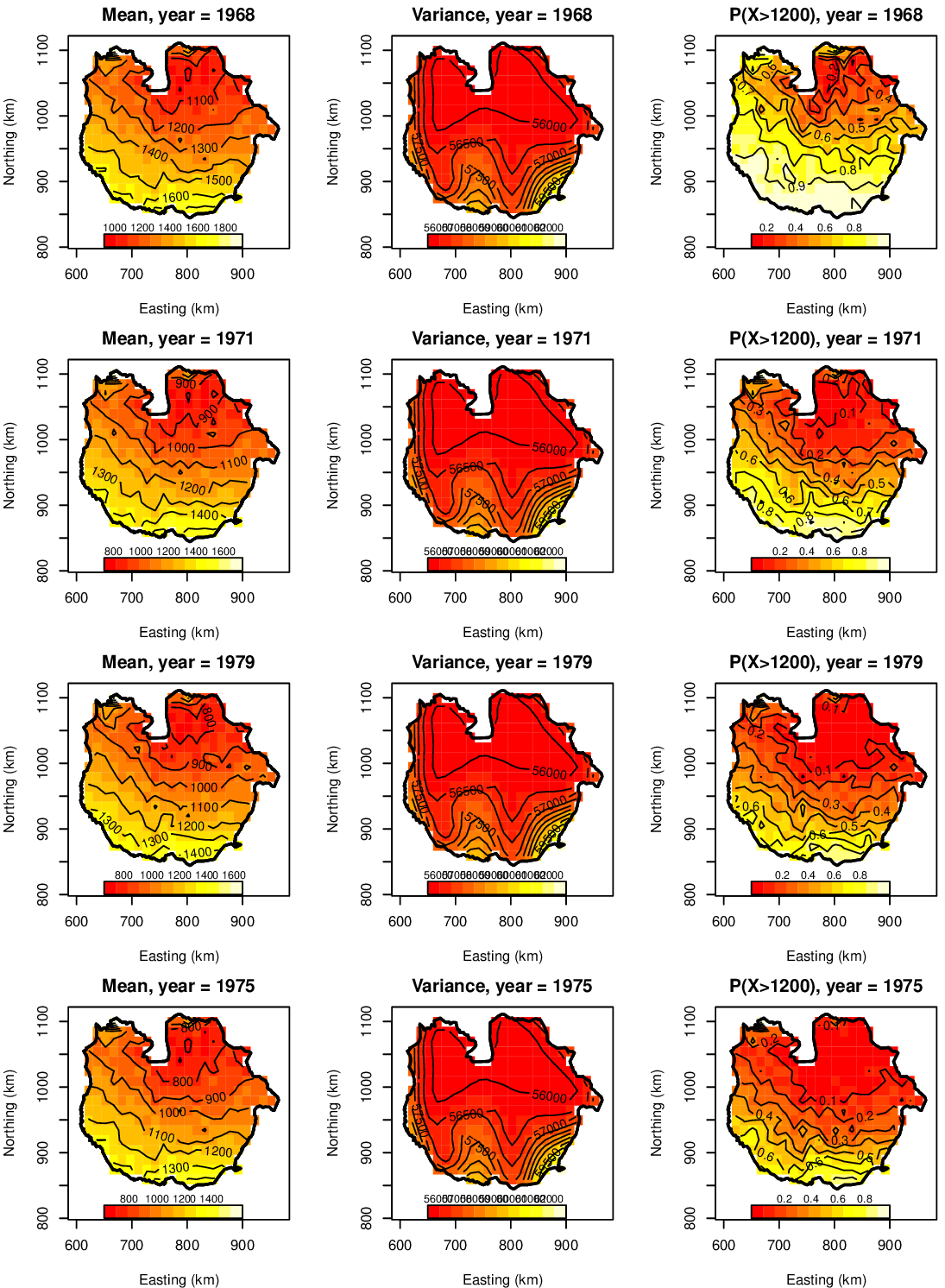}
    \caption{Posterior predictions for 1968, 1971, 1979, and 1975.}\label{fig_sp_3a}
\end{figure}

Figure \ref{fig_sp_3a} illustrates the posterior predictions for four selected years: 1968, 1971, 1979, and 1975. These years were chosen intentionally due to their observed mean annual rainfall, based on non-missing data, which decreases sequentially: approximately 1200 mm in 1968, 1100 mm in 1971, 1000 mm in 1979, and 900 mm in 1975. The maps present the posterior mean and variance of the predicted annual rainfall, as well as the probability of rainfall exceeding 1200 millimeters, plotted over a fixed grid. The color scheme uses red for low values (low precipitation, variance, and low probability of heavy rainfall), yellow for medium values, and white for high precipitation and probabilities.

In all four years, the posterior predictions identify regions of higher precipitation and a greater probability of rainfall exceeding 1200 millimeters in the central southern area and along the northern border of Gu{\'a}rico, where elevation is higher, relative to the rest of the state. While the spatial patterns remain consistent across the years, the overall mean rainfall varies, with 1968 being the wettest and 1979 the driest. The predictive maps also show an increasing prevalence of red areas over time, indicating a decline in predicted mean precipitation and a reduced probability of rainfall exceeding 1200 millimeters from 1968 to 1971, 1979, and 1975. This pattern aligns closely with the decreasing trend observed in the empirical data for these years.

\subsubsection{Remarks}

Overall, the coefficients of the predictors are consistent with the spatial and temporal patterns observed in the data. The predictions indicate a strong association between rainfall and the site's topographical features, such as whether it is situated on a plain or in the highlands, its proximity to forests or rivers, or the absence of such features. The model also captures a temporal trend effectively.

However, the model has some limitations. First, some model parameters exhibit large standard deviations, which may result from the grid search approach used for the correlation parameters \(\phi\) and \(\gamma^2\). This method limits the range of the posterior distribution that can be explored. Another possible reason is that the likelihood surface for this model is relatively flat over a wide range of values for \(\phi\) and \(\gamma^2\), leading to uncertainty in parameter estimation.

Additionally, the inclusion of both first- and second-order terms of the year in the trend component, while fitting the dataset well, may present challenges for long-term predictions. To address this, alternative approaches to modeling the temporal trend, such as incorporating it into the covariance matrix through auto-regression, could be considered for improved long-term forecasts.

Finally, computational constraints, due to the large covariance matrix (\(1280 \times 1280\)), resulted in a relatively short MCMC chain and fixing the order \(\nu\) in the Mat{\'e}rn correlation function. Extending the chain length and placing a prior on \(\nu\) could enhance the inference, providing more robust results and reducing potential biases.

\section{Bayesian Modeling of High-Order Markov Chains}

\cite{raftery1985} introduced the mixture transition distribution (MTD) model to facilitate efficient inference for time-homogeneous Markov chains (MCs) with high-order dependencies. Over the past three decades, frequentist methods have been developed to extend the MTD model and improve computational stability, although some challenges persist \citep{raftery1985}. This section focuses on presenting a Bayesian approach to analyzing the MTD model for time-homogeneous Markov chains.

\subsection{High-order Markov chains}

In high-order Markov chains, the state at time \( t \) depends on the previous \( \ell \) states, where \( \ell \geq 2 \). Let \( \{X_t\} \) denote an \( \ell \)-order \( m \)-state time-homogeneous Markov chain with a countable state space \( S \). The transition probabilities are defined as:
\[
\Pr(X_t = i_0 \mid X_{t-\ell} = i_\ell, \ldots, X_{t-1} = i_1) = q_{i_\ell \ldots i_0}, \quad i_t, \ldots, i_0 \in S\,.
\]
The full transition matrix of a high-order Markov chain is often sparse, containing many structural zeros. To simplify the presentation, we use a reduced form of the transition matrix. For instance, assuming \( \ell = 2 \) and \( m = 2 \), the transition matrix can be represented as:
\[
\mathbf{P} =
\begin{bmatrix}
p_{111} & p_{112} \\
p_{121} & p_{122} \\
p_{211} & p_{212} \\
p_{221} & p_{222}
\end{bmatrix}
\]

The maximum likelihood estimators (MLEs) for a fully parameterized \( \ell \)-order \( m \)-state Markov chain are given by:
\[
\hat{p}_{i_\ell \ldots i_0} = \frac{n_{i_\ell \ldots i_0}}{n_{i_\ell \ldots i_1+}}\,, \quad \text{where} \quad n_{i_\ell \ldots i_1+} = \sum_{i_0=1}^m n_{i_\ell \ldots i_0}\,.
\]
Here, \( n_{i_\ell \ldots i_0} \) represents the number of transitions from \( (X_{t-\ell} = i_\ell, \ldots, X_{t-1} = i_1) \) to \( X_t = i_0 \). In a fully parameterized \( \ell \)-order \( m \)-state Markov chain model, the number of free parameters is \( m^\ell (m-1) \).

\subsection{The MTD model}

The MTD model reduces the dimensionality of a fully parameterized model by decomposing the transition matrix \( \mathbf{P} \) into a linear combination of contributions from each of the past \( \ell \) states. The transition probability is expressed as:
\[
\Pr (X_t = i_0 \mid X_{t-\ell} = i_\ell, \ldots, X_{t-1} = i_1) = \sum_{g=1}^\ell \lambda_g \Pr(X_t = i_0 \mid X_{t-g} = i_g) = \sum_{g=1}^\ell \lambda_g q_{i_g i_0}\,, 
\]
where \( q_{i_g i_0} \) are elements of an \( m \times m \) transition matrix \( \mathbf{Q} = [q_{i_g i_0}] \), with each row forming a probability distribution. The vector \( \pmb{\lambda} = (\lambda_\ell, \ldots, \lambda_1) \) contains lag parameters representing the contribution from each lag.

\begin{figure}[!htb]
    \centering
    \includegraphics[scale=0.7]{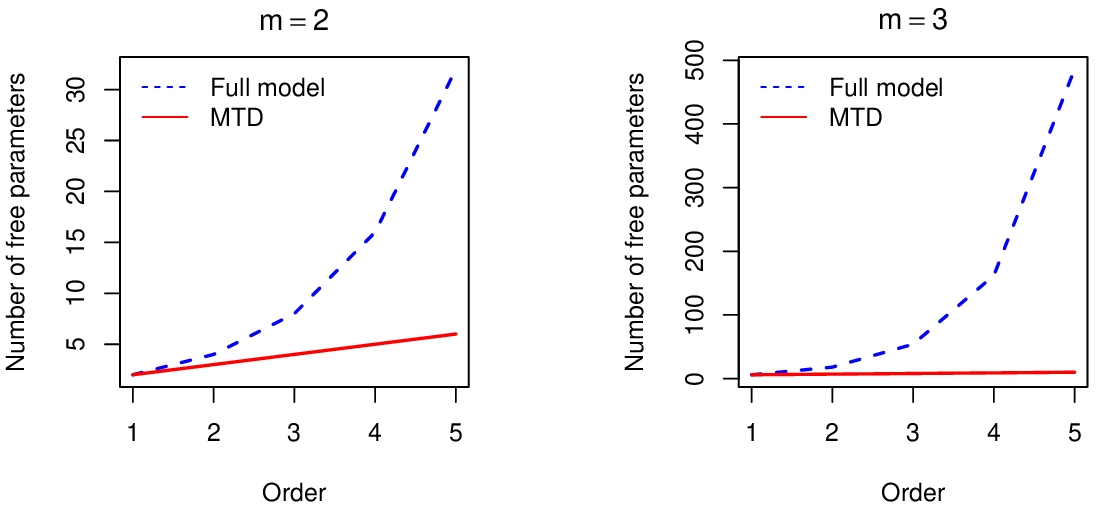}
    \caption{Number of free parameters under the full model and MTD.}\label{fig_homc_order}
\end{figure}

To ensure valid probabilities, the model imposes the following constraints:
\begin{equation}\label{eq_homc_1}
\sum_{g=1}^\ell \lambda_g = 1\,, \quad \lambda_g \geq 0\,, \quad g = 1, \ldots, \ell\,.
\end{equation}
Under the MTD model, the number of free parameters is \( m(m-1) + (\ell-1) \), significantly fewer than the \( m^\ell (m-1) \) parameters in the full model (see Figure \ref{fig_homc_order}). Frequentist estimation of the MTD model involves numerically maximizing the log-likelihood under the linear constraints defined in \eqref{eq_homc_1}, which can be computationally demanding. Although iterative algorithms have been proposed to enhance computational efficiency, they do not guarantee convergence to the global maximum of the log-likelihood \citep{berchtold2002mixture}.

\subsection{Modeling}

In contrast to frequentist estimation algorithms for MTD models, the Bayesian approach avoids the challenges associated with global maximization of the likelihood. For time-homogeneous MCs with a known order, MCMC algorithms can be applied to a model augmented with latent variables, enabling efficient parameter estimation. When the order is fixed but unknown, Bayesian methods leverage posterior model probabilities to perform order inference, providing a systematic approach to determining the most likely model complexity.

\subsubsection{Bayesian MTD models}

Let $\{X_t: t=1,\ldots,N+\ell\}$ represent a time-homogeneous $m$-state $\ell$-order MC with state space $S=\{1,\ldots,m\}$. Given the first $\ell$ observations, the likelihood of $\{X_t\}$ under the MTD model can be expressed as
\[
p(x_{\ell+1},\ldots,x_{\ell+N}\mid\pmb\lambda,\mathbf{Q},x_1,\ldots,x_\ell)
= \prod_{t=\ell+1}^{\ell+N}\left(\sum_{g=1}^\ell \lambda_g q_{x_{t-g},x_t}\right)\,,
\]
where $\pmb\lambda = (\lambda_1, \ldots, \lambda_\ell)$ is the vector of weights for the past $\ell$ lags, and $\mathbf{Q} = [q_{i,j}]$ is the $m \times m$ transition matrix for the MTD model.

From the above likelihood, it is apparent that the model resembles a mixture model. Introducing latent variables $\{w_t: t = \ell+1, \ldots, \ell+N\}$ as indicators for the mixture components allows the summation to be converted into a product. In this context, $\{w_t\}$ is a set of indicators specifying the mixture component from which each data point originates. The augmented likelihood then becomes
\[
p(x_{\ell+1}, \ldots, x_{\ell+N} \mid \pmb\lambda, \mathbf{Q}, w_{\ell+1}, \ldots, w_{\ell+N}, x_1, \ldots, x_\ell) = \prod_{t=\ell+1}^{\ell+N} \prod_{g=1}^\ell \left( \lambda_g q_{x_{t-g}, x_t} \right)^{I(w_t=g)}\,,
\]
where $I(\cdot)$ is the indicator function. This model can be efficiently estimated using MCMC methods. Finally, we assign independent Dirichlet prior distributions to $\pmb\lambda$ and each $\pmb{q}_i$:
\[
\pmb\lambda \sim \textsf{Dir}(b_1, \ldots, b_\ell)\,, \quad
\pmb q_i \sim \textsf{Dir}(a_{i,1}, \ldots, a_{i,m})\,, \quad i = 1, \ldots, m\,.
\]

\subsubsection{Computation}

Under the model given above, the posterior distribution is given by
\begin{align*}
p(x_{\ell+1},\ldots,x_{\ell+N},\,\pmb\lambda,\,\mathbf{Q},\,w_{\ell+1},\ldots,w_{\ell+N}\,|\,x_1,\ldots,x_\ell) 
&\propto 
\prod_{t=\ell+1}^{\ell+N}\prod_{g=1}^\ell \left(\lambda_g q_{x_{t-g},x_t}\right)^{I(w_t=g)} \\
&\qquad \cdot \prod_{i=1}^m\prod_{j=1}^m q_{ij}^{a_{ij}-1} \cdot \prod_{g=1}^\ell \lambda_g^{b_g-1}\,.
\end{align*}

\textit{MCMC algorithm for MCs with known order}

Based on the joint posterior, we run an MCMC algorithm to estimate the parameters using independent \(\mathrm{Dir}(1/2, \ldots, 1/2)\) priors for both \(\pmb\lambda\) and \(\pmb q_i\), for \(i = 1, \ldots, m\). The chain starts with 1,000 burn-in iterations, followed by 5,000 effective iterations. The algorithm is based on the following update steps:
\begin{enumerate}
    \item Sample \(w_t \sim \textsf{Mult}(\gamma_1, \ldots, \gamma_\ell)\), for \(t = \ell+1, \ldots, \ell+N\), where
    \[
    \gamma_g = \frac{\lambda_g\, q_{x_{t-g, t}}}{\sum_{r=1}^\ell \lambda_r\, q_{x_{t-r, t}}}, \quad g = 1, \ldots, \ell\,.
    \]

    \item Sample \(\pmb\lambda \sim \textsf{Dir}(b_1^\ast, \ldots, b_\ell^\ast)\), where
    \[
    b_g^\ast = b_g + \sum_{t=\ell+1}^{\ell+N} I(w_t = g), \quad g = 1, \ldots, \ell.
    \]

    \item Sample \(\pmb q_i \sim \textsf{Dir}(a_{i,1}^\ast, \ldots, a_{i,2}^\ast)\), where
    \[
    a_{i,j}^\ast = a_{i,j} + \sum_{t=\ell+1}^{\ell+N} I(x_{t-w_t} = i, \, x_t = j), \quad i, j = 1, \ldots, m.
    \]
\end{enumerate}

\textit{MCs with unknown order}

When the order of an MC is unknown, two main approaches can be considered. The first involves implementing a reversible jump MCMC algorithm, which allows for variable dimensions of the parameter vector \citep{richardson1997}. The second approach fits models of different orders and evaluates them using model selection criteria, such as the Bayes factor.

Here, we use posterior model probabilities for order selection \citep{scott2002} . Assume the chain \(\pmb x\) has an order \(\ell \in \{1, \ldots, L\}\), and let \(\pmb\phi = (\pmb\phi_1, \ldots, \pmb\phi_L)\) denote the vector of parameters, where \(\pmb\phi_\ell\) is the parameter vector for the \(\ell\)-th order MTD model. A Monte Carlo estimate of the posterior model probability, based on \(B\) draws from the posterior distribution, is given by \cite{scott2002}:
\[
p(\ell \mid \pmb x) = \int p(\ell \mid \pmb x, \pmb\phi) \, p(\pmb\phi \mid \pmb x)\,\text{d}\pmb\phi 
\approx \frac{1}{B} \sum_{j=1}^B p(\ell \mid \pmb x, \pmb\phi^{(j)}),
\]
where \(p(\ell \mid \pmb x, \pmb\phi^{(j)}) \propto p(\pmb x \mid \pmb\phi^{(j)}, \ell) \, p(\ell)\). Thus, the posterior model probability \(p(\ell \mid \pmb x)\) can be approximated by the mean of the posterior likelihood of the model, up to a proportionality constant.

\subsection{Illustration}

We implement the method on the infant death rate data assuming up to 5 orders of dependence and draw posterior inference on the order of the chain and the parameters using the aforementioned Bayesian methods. In this way, we use the provisional monthly data on infant death rate per 1,000 population in the United States between July 2011 and June 2014, as reported in the National Vital Statistics Reports, which consists of 36 consecutive months of data. The tables are publicly available on the CDC website, and the technical details are explained at \url{http://www.cdc.gov/nchs/data/nvsr/nvsr58/nvsr58_25.htm}.

\begin{figure}[!htb]
    \centering
    \includegraphics[scale=0.7]{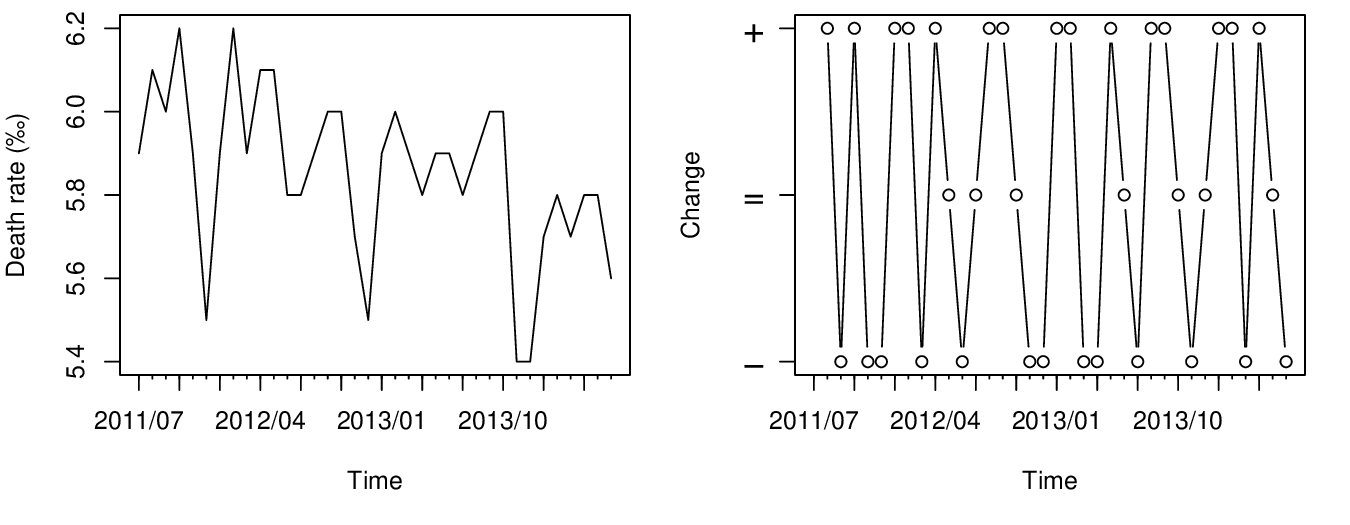}
    \caption{Monthly U.S infant death rate (per 1,000 population, left panel) and change in rate ($+$: increase, $=$: no change, $-$: decrease, right panel), from July 2011 to June 2014.}\label{fig_homc_1}
\end{figure}

Instead of using the raw data of the infant death rate (Figure \ref{fig_homc_1}, left panel), we analyze the change in the rate compared to the previous month. By assigning state 1 to a decrease, state 2 to no change, and state 3 to an increase, we transform the absolute rates into a three-state chain with state space \(S = \{1, 2, 3\}\). We assume up to 5 orders of dependence and take the first five observations as given throughout the analysis.

\subsubsection{Order of the chain}

We fit five Bayesian MTD models with orders \(\ell = 1, \ldots, 5\) and calculate the posterior model probability for each, as presented in Table \ref{tab_homc_postprob}. The second- and third-order models exhibit higher posterior model probabilities than the first-order model. Table \ref{tab_homc_postprob} highlights that the second-order MC (\(\ell = 2\)) is the most probable within the order space \(L = \{1, \ldots, 5\}\). Accordingly, we select the second-order model as the most suitable. For subsequent analyses, we adopt \(\ell = 2\) as the order of the MTD model.

\begin{table}[!htb]
\centering
\begin{tabular}{ccc}
\hline
\multirow{2}{*}{Order} & Mean Log-Posterior    & Posterior Model \\
                                  & Likelihood            & Probability \\ \hline
1                                 & -29.58                & 0.067        \\ 
2                                 & -27.17                & 0.738        \\ 
3                                 & -28.74                & 0.154        \\ 
4                                 & -30.25                & 0.034        \\ 
5                                 & -31.76                & 0.008        \\ \hline
\end{tabular}
\caption{Comparison of models based on posterior model probabilities.}
\label{tab_homc_postprob}
\end{table}

\subsubsection{Parameter estimation}

Table \ref{tab_homc_param} presents the summary statistics of the posterior samples for $\pmb\lambda$ and $q_{i,j}$'s. The substantially larger posterior mean of $\lambda_2$ compared to $\lambda_1$ suggests that $X_{t-2}$ has a stronger impact than $X_{t-1}$, which also explains the superior fit of the MTD model with $\ell=2$ over the first-order model.

\begin{table}[!htb]
    \centering
    \begin{tabular}{c c c c c c }
    \hline
    \multirow{2}{*}{Parameter} & \multirow{2}{*}{Prior Mean}&  \multicolumn{4}{c}{Posterior}\\\cline{3-6}
    & & Mean & SD & 2.5\% & 97.5\%\\
    \hline
    $\lambda_1$ & 0.50 &0.184&0.183&0.000&0.684\\
    $\lambda_2$ & 0.50 &0.816&0.183&0.316&1.000\\
    $q_{11}$ & 0.33&0.237&0.135&0.009&0.519\\
    $q_{12}$ & 0.33&0.716&0.142&0.419&0.965\\
    $q_{21}$ & 0.33&0.268&0.195&0.015&0.755\\
    $q_{22}$ & 0.33&0.284&0.179&0.007&0.675\\
    $q_{31}$ & 0.33&0.645&0.149&0.311&0.891\\
    $q_{32}$ & 0.33&0.232&0.127&0.044&0.525\\
    \hline\hline
    \end{tabular}
    \caption{Posterior distribution of $\pmb\lambda$ and $q_{i,j}$.}
    \label{tab_homc_param}
\end{table}

The posterior mean stationary distribution \(\pmb{\pi}\) represents the equilibrium probabilities of the system's state combinations, reflecting the long-term proportion of time the system spends in each state. In this context, the highest probabilities correspond to states \((1, 3)\), \((3, 1)\), and \((3, 3)\), with \(\pi_{13} = 0.153\), \(\pi_{31} = 0.150\), and \(\pi_{33} = 0.155\), indicating these states dominate the system's dynamics. Conversely, states such as \((2, 2)\) and \((2, 1)\), with stationary probabilities of \(\pi_{22} = 0.073\) and \(\pi_{21} = 0.083\), are visited less frequently, highlighting their lower significance in the system's behavior.

\begin{figure}[!htb]
    \centering
    \includegraphics[scale=0.7]{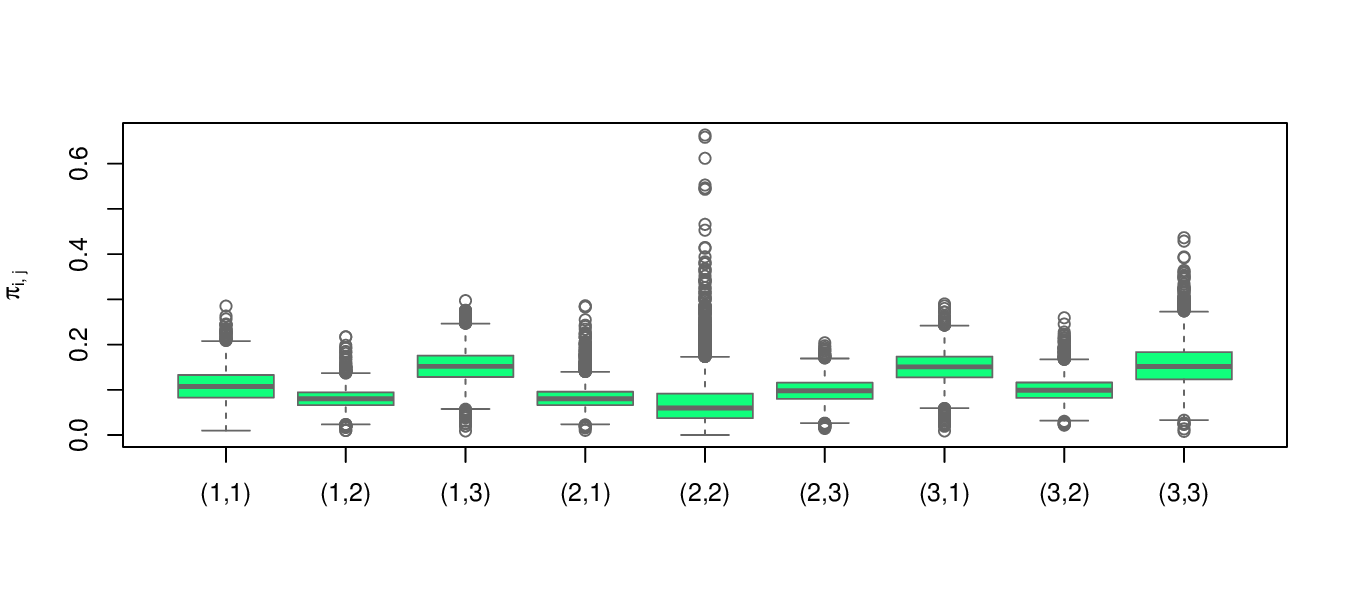}
    \caption{Box plots of posterior samples of \(\pmb{\pi}\).}\label{fig_homc_pi}
\end{figure}

Figure \ref{fig_homc_pi} displays box plots corresponding to the posterior samples of \(\pmb{\pi}\), highlighting a considerable number of samples that fall outside the 1.5 interquartile range for $\pi_{(2,2)}$. This is largely attributed to the small size of the original dataset (30 data points, excluding the first five observations treated as given) and the lack of observed transitions from state 2 to state 2. Expanding the sample size could mitigate the spread of the distributions and enhance the precision of the estimates.

\subsubsection{Model assessment}

We employ the posterior predictive loss criterion proposed by \cite{gelfand1998} to evaluate the model's performance. This criterion selects the model \( \mathcal{M} \) that minimizes:
\[
D_r(\mathcal{M}) = P^{\mathcal{M}} + \frac{r}{r+1} G^{\mathcal{M}}
\]
where
\[
P^{\mathcal{M}} = \sum_{t=1}^n \textsf{Var}^{\mathcal{M}}(\hat{y}_t), \quad 
G^{\mathcal{M}} = \sum_{t=1}^n \left( y_t - \textsf{E}^{\mathcal{M}}(\hat{y}_t) \right)^2,
\]
and \( r \geq 0 \). Here, \( \textsf{E}^{\mathcal{M}}(\hat{y}_t) \) and \( \textsf{Var}^{\mathcal{M}}(\hat{y}_t) \) denote the mean and variance of the posterior predictive distribution at time \( t \) under model \( \mathcal{M} \). The criterion consists of two components: The penalty term \( P^{\mathcal{M}} \), which accounts for model complexity, and the goodness-of-fit term \( G^{\mathcal{M}} \), weighted by \( \frac{r}{r+1} \). This balance ensures a trade-off between model parsimony and fit to the data.

The posterior predictive loss criterion described above is used to compare the estimated \(\ell\)-th order MTD model \(M_\ell\) with the fully parameterized model \(M_0\). For the fully parameterized model \(M_0\), the \(m^\ell \times m\) transition matrix \(\mathbf{P}\) is also estimated within a Bayesian framework. In this approach, each row of transition probabilities in \(\mathbf{P}\) is assigned an independent Dirichlet\((1/2, \ldots, 1/2)\) prior, reflecting a weakly informative prior distribution that encourages balance while allowing flexibility in the transition probabilities. This enables a fair comparison between the simpler \(M_\ell\) model and the more complex \(M_0\), taking into account both model fit and complexity.

We fit the fully parameterized second-order Markov chain model (\(M_0\)) and compare its posterior predictive loss to that of the second-order MTD model (\(M_2\)), with the results summarized in Table \ref{tab_homc_PGD}. The comparison reveals that \(M_2\) has lower values for both \(D_1\) and \(D_\infty\), indicating a preference for the second-order MTD model over the fully parameterized model in this case. Analyzing the \(P\) and \(G\) terms shows that \(M_2\) achieves a comparable fit to the data as \(M_0\), while being more parsimonious. This highlights the advantage of the MTD model in balancing model fit and complexity.

\begin{table}[!htb]
    \centering
    \begin{tabular}{c c c c c c}
        \hline
        Model  & Parameters & ~~~P~~~ & ~~~G~~~ & ~~~ D$_1$ ~~~ & ~~~ D$_\infty$ ~~~ \\
        \hline
        M$_0$ & 18 & 23.21 & 20.38 & 33.4 & 43.59 \\
        M$_2$ &  7 & 21.46 & 20.88 & 31.9 & 42.34 \\
        \hline
    \end{tabular}
    \caption{Posterior predictive loss of fully parameterized second-order Markov chain model (\(M_0\)) and second-order MTD model (\(M_2\)).}
    \label{tab_homc_PGD}
\end{table}

\subsubsection{Prediction}

Given the weight vector \(\lambda\) and the MTD transition matrix \(\mathbf{Q}\), the \(\ell\)-order transition matrix \(\mathbf{P}\) can be reconstructed as follows:
\[
\mathbf{P}_{i_\ell\ldots i_1i_0} = P(X_t = i_0 \mid X_{t-\ell} = i_\ell, \ldots, X_{t-1} = i_1) = \sum_{g=1}^\ell \lambda_g q_{i_g i_0}, \quad i_\ell, \ldots, i_0 \in S,
\]
where \(q_{i_g i_0}\) are the elements of the MTD transition matrix \(\mathbf{Q} = [q_{i_g i_0}]\). This formulation leverages the structure of the MTD model, which simplifies the high-dimensional \(\ell\)-order Markov chain into a weighted combination of simpler transitions.

To predict future sample paths, new observations can be generated sequentially, conditioned on the observed data. By averaging over the posterior sample draws obtained from the MCMC algorithm, the posterior predictive probability of a specific state at each time point can be calculated. This approach provides a probabilistic framework for forecasting, integrating uncertainty from the posterior distribution into the predictions.

\begin{figure}[!htb]
    \centering
    \includegraphics[scale=0.7]{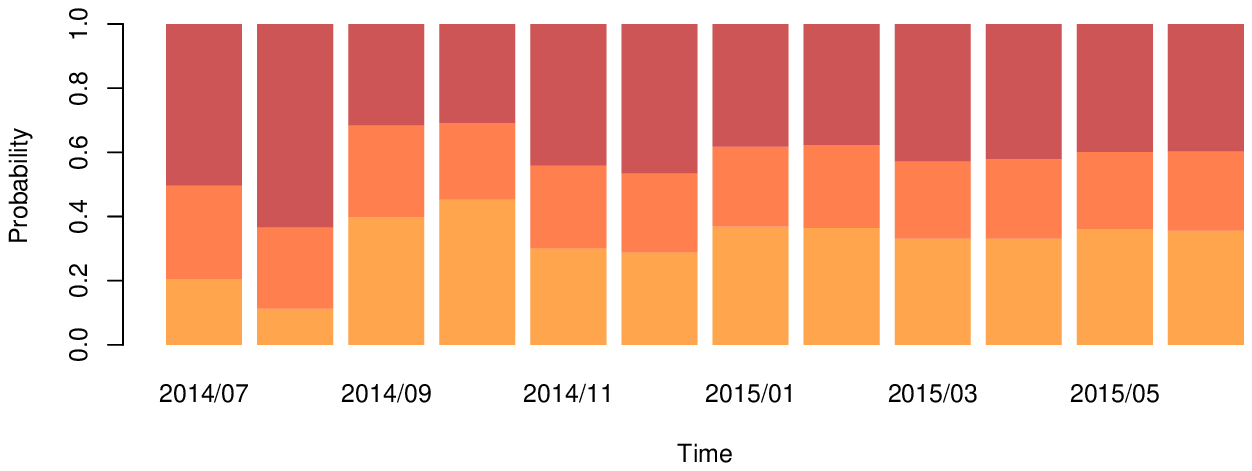}
    \caption{Bar chart of posterior predicted probability of states for the next 12 months (1: Tan, 2: Coral, 3: Red).}\label{fig_homc_pred}
\end{figure}

For each posterior sample, we reconstruct the full transition matrix \(\mathbf{P}\) and use it to predict the relative change in the infant death rate over the subsequent 12 months. The results are summarized in Figure \ref{fig_homc_pred}, which illustrates the predicted probabilities for each state in the following year. In this figure, the states represent the relative changes: tan indicates a decrease, coral represents no change, and red signifies an increase compared to the previous month. These probabilities, derived from the posterior samples, provide insights into the expected dynamics of the infant death rate over time, accounting for uncertainty in the model estimates.

\subsubsection{Remarks}

We present a Bayesian framework for analyzing Mixture Transition Distribution (MTD) models in high-order Markov chains. By assessing posterior model probabilities, we determine the optimal order of dependence in the data, formulating this task as a Bayesian model selection problem where the model with the highest posterior probability serves as the selection criterion. The chosen MTD model is then compared to the fully parameterized Markov model, demonstrating superior performance in balancing goodness-of-fit and model parsimony. Furthermore, we generate near-future predictions, illustrating the practical utility of the MTD approach. Our results indicate that Bayesian MTD models can be effectively implemented using MCMC methods and provide strong performance relative to the full Markov model, making them a compelling choice for modeling high-order dependencies.

\section{A Variational Algorithm for Dirichlet Process Mixtures}

The mean field variational inference method provides a framework for approximating complex posterior distributions by reformulating the problem as an optimization task. In Bayesian inference, the posterior distribution \( p(\pmb\theta \mid \pmb y) \) often involves computationally intractable integrals, especially in high-dimensional spaces. Variational methods \citep{jordan1999} address this challenge by replacing the exact posterior with an approximate distribution \( q_{\pmb\eta}(\pmb\theta) \), parameterized by variational parameters \(\pmb\eta\), and optimizing the fit between the two distributions. The goal is to minimize the Kullback-Leibler (KL) divergence, which quantifies the difference between \( q_{\pmb\eta}(\pmb\theta) \) and \( p(\pmb\theta \mid \pmb y) \). This divergence is defined as 
$$
    KL(p \| q) = \int_\Theta \log \frac{q_{\pmb\eta}(\pmb\theta)}{p(\pmb\theta \mid \pmb y)} \, q_{\pmb\eta}(\pmb\theta)\, \text{d}\pmb\theta\,,
$$
where \( \pmb\theta \) represents latent variables within a parameter space \( \Theta \). By minimizing this divergence, the variational approach ensures that \( q_{\pmb\eta}(\pmb\theta) \) closely approximates the true posterior \citep{bishop2006}.

The mean field approach, foundational to variational inference methods simplifies the problem by assuming that the posterior distribution can be factorized into independent components. Specifically, the joint posterior \( p(\pmb\theta \mid \pmb y) \) is approximated as 
$$
    q_{\pmb\eta}(\pmb\theta) = \prod_{k=1}^K q_{\eta_k}(\theta_k)\,, 
$$
where each \( q_{\pmb\eta_k}(\pmb\theta_k) \) represents the distribution of the \( k \)-th latent variable, governed by its variational parameters. This factorization significantly reduces the computational complexity of working with the full posterior, allowing efficient optimization. Early implementations of this method in probabilistic graphical models demonstrated its scalability and practicality for high-dimensional inference \citep{koller2009}.

The algorithm iteratively updates the variational parameters \( \eta_k \) to refine the approximation. For each latent variable, the full conditional posterior \( p(\theta_k \mid \theta_{-k}, \pmb y) \) is used as a reference. In many practical settings, this conditional posterior belongs to the exponential family, and its form can be written as
$$
    p(\pmb\theta_k \mid \pmb\theta_{-k}, \pmb y) \propto \exp\left\{\sum_{\ell=1}^L g_{k,\ell}(\pmb\theta_{-k}, \pmb y) \, t_\ell(\pmb\theta_k) - h(\pmb\theta_k)\right\}\,.
$$
Here, \( t_\ell(\pmb\theta_k) \) represents the sufficient statistics of the conditional distribution, \( g_{k,\ell}(\pmb\theta_{-k}, \pmb y) \) are functions capturing the dependencies on other latent variables \( \pmb\theta_{-k} \) and the observed data \( \pmb y \), and \( h(\pmb\theta_k) \) is a normalizing term. This approach follows from the exponential family framework provided in \cite{blei2017}.

The optimization step involves deriving the variational distribution \( q_{\eta_k}(\theta_k) \) that best approximates \( p(\theta_k \mid \theta_{-k}, \pmb y) \). This is achieved by setting the parameters \( \eta_{k,\ell} \) of \( q_{\eta_k}(\theta_k) \) to the expected value of the corresponding terms in the conditional posterior: \( \eta_{k,\ell} = \textsf{E}_{q_{-k}}[g_{k,\ell}(\theta_{-k}, \pmb y)] \). Here, the expectation is taken with respect to \( q_{-k} \), the variational distribution over all latent variables except \( \theta_k \). This expectation ensures that the parameters of \( q_{\eta_k}(\theta_k) \) incorporate information from the rest of the model \citep{hoffman2013}.

The iterative process of updating \( \eta_k \) for each latent variable continues until the KL divergence converges to a minimum, providing an optimal approximation for the posterior. This approach, by leveraging factorization and conditional independence, allows for scalable inference in high-dimensional models. Mean field variational inference thus offers a powerful and computationally efficient alternative to exact Bayesian methods, striking a balance between accuracy and feasibility in complex problems.

\subsection{Dirichlet process mixture model}

In a Dirichlet Process Mixture Model (DPMM; e.g., \citealt{rodriguez2013nonparametric}, \citealt{muller2013bayesian}, \citealt{muller2018nonparametric}), the Dirichlet Process (DP; \citealt{ferguson1973bayesian} and \citealt{ferguson1974prior}) serves as a prior for the mixing distribution \( G \), enabling an unknown and potentially infinite number of mixture components. This approach is particularly effective for analyzing data with complex structures or when the number of underlying subgroups (clusters) is unknown. The model is formulated as:
\[
F(\cdot \mid G) = \int K(\cdot \mid \theta) \, \text{d}G(\theta)\,,
\]
where \( G \sim DP(\alpha, G_0) \) indicates that \( G \) is drawn from a DP with concentration parameter \( \alpha \) and base measure \( G_0 \), and \( K(\cdot \mid \theta) \) is a parametric distribution defined by parameters \( \theta \). Each \( \theta \) specifies the parameters for an individual component of the mixture. In this way, the corresponding mixture density or probability mass function in the DPMM can be expressed as:
\[
f(\cdot \mid G) = \int k(\cdot \mid \theta) \, \textsf{d}G(\theta)\,,
\]
where \( k(\cdot \mid \theta) \) denotes the density (or probability mass function) of \( K(\cdot \mid \theta) \). Since \( G \) is random, the mixture density \( f(\cdot \mid G) \) and the cumulative distribution \( F(\cdot \mid G) \) are also random.

In a DPMM, a DP prior is placed on \( G \), the distribution of parameters \( \theta \). The DP ensures that \( G \) is a discrete distribution, potentially consisting of infinitely many "atoms" or components. Thus, data points \( y_i \) are generated in two steps: First, a parameter \( \theta_i \) is drawn from \( G \), and second, \( y_i \) is sampled from \( K(\cdot \mid \theta_i) \), a distribution governed by \( \theta_i \). Since \( G \) is discrete, many of the \( \theta_i \) values will coincide, naturally forming groupings or clusters among the \( y_i \) values. This characteristic enables the DPMM to infer clusters in the data adaptively, even without prior knowledge of the number of components.

\subsubsection{Alternative formulation}

Using the constructive definition of the DP \citep{sethuraman1994constructive}, a random distribution \( G \) drawn from a \( DP(\alpha, G_0) \) can be represented as:
\[
G = \sum_{\ell=1}^{\infty} p_{\ell}\, \delta_{Z_{\ell}}\,,
\]
where \( \delta_{a} \) denotes a Dirac delta function centered at \( a \). This representation, known as the stick-breaking process, provides an intuitive way to interpret \( G \) as a countable mixture of point masses, each located at parameters \( Z_{\ell} \) with corresponding weights \( p_{\ell} \).

Under this framework, the probability model \( f(\cdot \mid G) \) can be expressed as a countable mixture of parametric densities:
\[
f(\cdot \mid G) = \sum_{\ell=1}^{\infty} p_{\ell} \, k(\cdot \mid Z_{\ell})\,,
\]
where \( k(\cdot \mid Z_{\ell}) \) is a parametric density function parameterized by \( Z_{\ell} \), and \( p_{\ell} \) are the weights generated by the stick-breaking process. The construction begins with the first weight \( p_1 = v_1 \), and subsequent weights are defined recursively as \( p_{\ell} = v_{\ell} \prod_{r=1}^{\ell-1} (1 - v_r) \), where \( v_r \sim \textsf{Beta}(1, \alpha) \) for \( r = 1, 2, \ldots \). The stick-breaking process guarantees that the weights \( p_1, p_2, \ldots \) sum to 1, ensuring a valid probability distribution over the infinite components.

The locations \( Z_{\ell} \) are independently sampled from the base measure \( G_0 \), which serves as the prior over the parameter space and encodes prior knowledge about the data-generating process. Importantly, the sequences \( \{v_r : r = 1, 2, \ldots\} \) and \( \{Z_{\ell} : \ell = 1, 2, \ldots\} \) are independent, ensuring that the generation of weights and the selection of component locations are decoupled. This separation simplifies the modeling and enhances the interpretability of the DP-based mixture.

\subsubsection{Hierarchical formulation}

Typically, DPMMs are employed to model data as follows:
\[
y_i \mid G, \phi \overset{\text{iid}}{\sim} f(\cdot \mid G, \phi) = \int k(\cdot \mid \theta, \phi) \, \text{d}G(\theta)\,, \quad i = 1, \ldots, n\,,
\]
where \( G \sim \textsf{DP}(\alpha, G_0) \) is a random distribution drawn from a DP with concentration parameter \( \alpha \) and base measure \( G_0 \). Then, a parametric prior \( p(\phi) \) is placed on the global parameter \( \phi \), and hyperpriors may be specified for \( \alpha \) or the parameters \( \pmb\vartheta \) governing \( G_0 = G_0(\cdot \mid \pmb\vartheta) \), as follows:
\begin{enumerate}
\item Observation model: Each \( y_i \) is conditionally independent, given its associated latent parameter \( \theta_i \) and the global parameter \( \phi \):
   \[
   y_i \mid \theta_i, \phi \sim k(y_i \mid \theta_i, \phi), \quad i = 1, \ldots, n\,.
   \]

\item Latent mixing parameters: The latent variables \( \theta_i \) are independently drawn from the random distribution \( G \):
   \[
   \theta_i \mid G \sim G, \quad i = 1, \ldots, n\,.
   \]

\item DP prior: The mixing distribution \( G \) is itself drawn from a Dirichlet Process:
   \[
   G \mid \alpha, \pmb\vartheta \sim \textsf{DP}(\alpha, G_0(\cdot \mid \pmb\vartheta))\,,
   \]
   where \( \alpha \) controls the concentration of \( G \), and \( G_0 \) acts as the base measure, parameterized by \( \psi \).

\item Priors on model parameters: Independent priors are placed on the global parameter \( \phi \), the concentration parameter \( \alpha \), and the base measure parameter \(\pmb\vartheta \):
   \[
   \phi, \alpha, \pmb\vartheta \sim p(\phi) \, p(\alpha) \, p(\pmb\vartheta)\,.
   \]
\end{enumerate}

In the first step of the previous formulation, the mixing parameters \( \theta_i \) can be replaced with configuration variables \( L_1, L_2, \ldots \). Each \( L_i \) is defined such that \( L_i = \ell \) if and only if \( \theta_i = Z_\ell \), for \( i = 1, 2, \ldots \) and \( \ell = 1, 2, \ldots \). This representation simplifies the model by associating each observation \( i \) with one of the distinct components indexed by \( \ell \), offering a more concise framework for managing clustering assignments. Therefore, observation model can be rewritten as:
\begin{align*}
    y_i \mid L_i, Z_{L_i}, \phi &\overset{\text{ind}}{\sim} k(y_i \mid Z_{L_i}, \phi) \\
    L_i \mid \pmb p &\overset{\text{iid}}{\sim} \sum_{\ell=1}^N p_\ell \, \delta_\ell(L_i) \\
    Z_\ell \mid \pmb\vartheta &\overset{\text{iid}}{\sim} G_0(\cdot \mid \pmb\vartheta)\,,
\end{align*}
 for \( i = 1, 2, \ldots \) and \( \ell = 1, 2, \ldots \).

\subsection{Illustration}

We propose a mean field variational algorithm to perform inference on a DPMM, where \( y_i \mid L_i, Z_{L_i}, \phi \overset{\text{ind}}{\sim} \textsf{N}(y_i \mid Z_{L_i}, 1/\phi) \), for \( i = 1, \ldots, n \). The model assumes a parametric prior \( \phi \sim \textsf{G}(a_\phi, b_\phi) \) on the precision parameter, a fixed concentration parameter \( \alpha = 1 \), and a baseline distribution \( G_0 = \textsf{N}(\psi, 1/\nu) \).

Here, \( a_\phi \), \( b_\phi \), \( \psi \), and \( \nu \) are treated as hyperparameters. While this formulation considers \( \psi \), \( \nu \), and \( \alpha \) as fixed, they can also be assigned additional priors to incorporate further uncertainty into the model. This flexibility allows for a fully Bayesian treatment when desired, enabling richer inference while maintaining computational tractability. In this case, the posterior distribution is given by:
\begin{align}
p(\pmb{Z}, \pmb{L}, \pmb{v}, \phi \mid \pmb{y}) 
&\propto \prod_{\ell=1}^\infty v_\ell^{1-1}(1-v_\ell)^{\alpha-1} 
\prod_{\ell=1}^\infty \exp\left\{-\frac{1}{2} \nu (Z_\ell - \psi)^2 \right\} \nonumber \\
&\quad \cdot \prod_{\ell=1}^\infty \phi^{\frac{1}{2} \sum_{i=1}^n I(L_i = \ell)} 
\exp\left\{-\frac{1}{2} \sum_{i=1}^n I(L_i = \ell) \phi (y_i - Z_\ell)^2 \right\} \nonumber \\
&\quad \cdot \prod_{\ell=1}^\infty\prod_{i=1}^n  v_\ell^{I(\ell = L_i)} 
(1 - v_\ell)^{I(\ell < L_i)} \cdot \phi^{a_\phi - 1} 
\exp\left\{-b_\phi \phi \right\}\,,
\end{align}
where $I(\cdot)$ is the indicator function.

\subsubsection{Variational algorithm}

The variational distribution for this model is designed with a truncation level \( N \), which approximates the infinite-dimensional stick-breaking representation of the DP. To simplify the representation, boundary conditions are imposed such that \( q(v_N = 1) = 1 \) and \( q(v_\ell = 0) = 1 \) for \( \ell > N \), effectively truncating the stick-breaking process beyond the \( N \)-th component.

The variational distribution is formulated as a factorized family of distributions, given by:  
\[
q(\phi, \pmb{v}, \pmb{Z}, \pmb{L}) = 
q_{\xi}(\phi) \cdot \prod_{\ell=1}^{N-1} q_{\gamma_\ell}(v_\ell) 
\cdot \prod_{\ell=1}^N q_{\eta_\ell}(Z_\ell) 
\cdot \prod_{i=1}^n q_{\varpi_i}(L_i).
\]
Under this approximation, the precision parameter \( \phi \) follows a Gamma distribution, \( q_\xi(\phi) = \textsf{G}(\xi_1, \xi_2) \), the stick-breaking weights \( v_\ell \) follow a Beta distribution, \( q_{\gamma_\ell}(v_\ell) = \textsf{Beta}(\gamma_{\ell,1}, \gamma_{\ell,2}) \), the component-specific parameters \( Z_\ell \) follow a Normal distribution, \( q_{\eta_\ell}(Z_\ell) = \textsf{N}(\eta_{\ell,1}, \eta_{\ell,2}) \), and finally, the latent cluster assignments \( L_i \) follow a Multinomial distribution, \( q_{\varpi_i}(L_i) = \textsf{Mult}(\varpi_i) \). Here, \( \xi_1 \), \( \xi_2 \), \( \gamma_{\ell,1} \), \( \gamma_{\ell,2} \), \( \eta_{\ell,1} \), \( \eta_{\ell,2} \), and \( \varpi_i \) denote the variational parameters.

This variational distribution leverages conjugate forms for the posterior approximations, enabling efficient optimization of the variational parameters during inference. The independence assumption and the choice of conjugate distributions make this approach computationally feasible, even for large datasets and high truncation levels \( N \).

The variational update rules for the parameters are outlined below:
\begin{enumerate}
    \item Precision Parameter:  
    The posterior conditional density is given by:
    \begin{align*}
        p(\phi \mid \ldots) &\propto \phi^{a_\phi + \frac{1}{2} \sum_{\ell=1}^N \sum_{i=1}^n I(L_i = \ell) - 1} \\ 
        &\quad\cdot \exp\Bigg\{-\phi \Bigg(b_\phi + \frac{1}{2} \sum_{\ell=1}^N \sum_{i=1}^n I(L_i = \ell) (y_i - Z_\ell)^2 \Bigg)\Bigg\}.
    \end{align*}
    The variational parameters are updated as:
    \begin{align*}
        \xi_1 &= a_\phi + \frac{1}{2} \sum_{i=1}^n \sum_{\ell=1}^N \textsf{E}[I(L_i = \ell)] = a_\phi + \frac{n}{2}, \\
        \xi_2 &= b_\phi + \frac{1}{2} \sum_{\ell=1}^N \sum_{i=1}^n \textsf{E}[I(L_i = \ell)] \cdot \Big[y_i^2 - 2y_i \textsf{E}(Z_\ell) + \textsf{E}(Z_\ell)^2 + \textsf{Var}(Z_\ell)\Big].
    \end{align*}
    Substituting \(\textsf{E}[I(L_i = \ell)] = \varpi_{i,\ell}\), \(\textsf{E}(Z_\ell) = \eta_{\ell,1}\), and \(\textsf{Var}(Z_\ell) = \eta_{\ell,2}\), we have:
    \begin{align*}
        \xi_2 &= b_\phi + \frac{1}{2} \sum_{\ell=1}^N \sum_{i=1}^n \varpi_{i,\ell} \Big(y_i^2 - 2y_i \eta_{\ell,1} + \eta_{\ell,1}^2 + \eta_{\ell,2}\Big).
    \end{align*}

    \item Stick-breaking weights: 
    The posterior conditional density is given by:
    \begin{align*}
        p(v_\ell \mid \ldots) &\propto v_\ell^{1 + \sum_{i=1}^n I(L_i = \ell) - 1} (1 - v_\ell)^{\alpha + \sum_{i=1}^n I(L_i > \ell) - 1}.
    \end{align*}
    The variational parameters are updated as:
    \begin{align*}
        \gamma_{\ell,1} &= 1 + \sum_{i=1}^n \textsf{E}[I(L_i = \ell)] = 1 + \sum_{i=1}^n \varpi_{i,\ell}, \\
        \gamma_{\ell,2} &= \alpha + \sum_{i=1}^n \textsf{E}[I(L_i > \ell)] = \alpha + \sum_{i=1}^n \sum_{j=\ell+1}^N \varpi_{i,j}.
    \end{align*}

    \item Component Parameters:  
    The posterior conditional density is:
    \begin{align*}
        p(Z_\ell \mid \ldots) &\propto \exp\Bigg(-\frac{1}{2} \Bigg\{ \Big(\nu + \sum_{i=1}^n I(L_i = \ell) \phi\Big) Z_\ell^2 \nonumber \\
        &\qquad\qquad\qquad\qquad - 2\Big(\nu\psi + \sum_{i=1}^n I(L_i = \ell) \phi y_i\Big) Z_\ell \Bigg\}\Bigg).
    \end{align*}
    The variational parameters are updated as:
    \begin{align*}
        \eta_{\ell,1} &= \eta_{\ell,2} \cdot \Bigg(\nu\psi + \sum_{i=1}^n \textsf{E}[I(L_i = \ell)] \cdot \textsf{E}(\phi) \cdot y_i\Bigg), \\
        \eta_{\ell,2} &= \left(\nu + \sum_{i=1}^n \textsf{E}[I(L_i = \ell)] \cdot \textsf{E}(\phi)\right)^{-1}.
    \end{align*}
    Substituting \(\textsf{E}[I(L_i = \ell)] = \varpi_{i,\ell}\) and \(\textsf{E}(\phi) = \xi_1 / \xi_2\), we have:
    \begin{align*}
        \eta_{\ell,1} &= \left(\nu + \frac{\xi_1}{\xi_2} \sum_{i=1}^n \varpi_{i,\ell}\right)^{-1} \Bigg(\nu\psi + \frac{\xi_1}{\xi_2} \sum_{i=1}^n \varpi_{i,\ell} y_i\Bigg), \\
        \eta_{\ell,2} &= \left(\nu + \frac{\xi_1}{\xi_2} \sum_{i=1}^n \varpi_{i,\ell}\right)^{-1}.
    \end{align*}

    \item Cluster Assignments:  
    The cluster probabilities are given by:
    \begin{align*}
        p(L_i = \ell \mid \ldots) &\propto v_\ell \prod_{j=1}^{\ell-1} (1 - v_j) \cdot \phi^{\frac{1}{2}} \exp\Bigg(-\frac{1}{2} \phi \Big(y_i - Z_\ell\Big)^2\Bigg).
    \end{align*}
    Taking expectations over the variational distributions, \(\varpi_{i,\ell}\) is updated as:
    \begin{align*}
        \varpi_{i,\ell} &= \textsf{E}[\log(v_\ell)] + \sum_{j=1}^{\ell-1} \textsf{E}[\log(1 - v_j)] + \frac{1}{2} \textsf{E}[\log(\phi)] \nonumber \\
        &\qquad\qquad\qquad - \frac{1}{2} \textsf{E}(\phi) \Big[y_i^2 - 2y_i \textsf{E}(Z_\ell) + \textsf{E}(Z_\ell)^2 + \textsf{Var}(Z_\ell)\Big].
    \end{align*}
    Using \(\textsf{E}[\log(v_\ell)] = \Gamma'(\gamma_{\ell,1}) - \Gamma'(\gamma_{\ell,1} + \gamma_{\ell,2})\), and similar expansions:
    \begin{align*}
        \varpi_{i,\ell} &= \Gamma'(\gamma_{\ell,1}) - \Gamma'(\gamma_{\ell,1} + \gamma_{\ell,2}) + \sum_{j=1}^{\ell-1} \Big[\Gamma'(\gamma_{j,2}) - \Gamma'(\gamma_{j,1} + \gamma_{j,2})\Big] \nonumber \\
        &\quad\qquad\qquad + \frac{1}{2} \Big[\Gamma'(\xi_1) - \log(\xi_2)\Big] - \frac{\xi_1}{2\xi_2} \Big[y_i^2 - 2y_i \eta_{\ell,1} + \eta_{\ell,1}^2 + \eta_{\ell,2}\Big].
    \end{align*}
\end{enumerate}

The convergence of the variational algorithm is monitored using a threshold for the infinity norm of the variational parameter differences, which is defined as the maximum absolute difference between corresponding elements of the variational parameter vector across successive iterations of the algorithm. This criterion ensures that the updates to the parameters stabilize, allowing the algorithm to reach a reliable approximation. A threshold of \( 10^{-5} \) is employed to signify a high level of precision in the iterative optimization process, providing a clear indication of when the algorithm has effectively converged.

The predictive density quantifies the likelihood of a new observation \( y_{n+1} \) given the observed data \( \pmb y \). It is approximated by summing over all possible components of the model, weighted by their probabilities under the variational posterior:
\[
p(y_{n+1} \mid \pmb y) \approx \sum_{\ell=1}^N \textsf{E}_q\{p_\ell\} \, \textsf{E}_q\left(\textsf{N}(y_{n+1} \mid Z_\ell, \phi)\right).
\]
Here, \( \textsf{E}_q\{p_\ell\} \) represents the expected contribution of each component, derived from the stick-breaking construction:
\[
\textsf{E}_q\{p_\ell\} = \frac{\gamma_{\ell,1}}{\gamma_{\ell,1} + \gamma_{\ell,2}}
\left(\prod_{j=1}^{\ell-1} \frac{\gamma_{j,2}}{\gamma_{j,1} + \gamma_{j,2}}\right).
\]
This formulation incorporates the uncertainty in the component assignments, providing a probabilistic forecast for new data while accounting for the underlying structure of the model.

The variational bound of the log marginal likelihood serves as a measure of how well the variational distribution approximates the true posterior. This bound is expressed as:
\begin{align*}
    \log p(\pmb y \mid \pmb\theta, \pmb\eta) &\geq \textsf{E}_q[\log p(\phi \mid a_\phi, b_\phi)] 
    + \textsf{E}_q[\log p(\pmb v \mid \alpha)] \nonumber \\
    &\quad + \sum_{\ell=1}^N \textsf{E}_q[\log p(Z_\ell \mid \psi, \nu)] \nonumber \\
    &\quad + \sum_{i=1}^n \Big( \textsf{E}_q[\log p(L_i \mid \pmb v)] + \textsf{E}_q[\log p(y_i \mid Z_{L_i}, \phi)] \Big) \nonumber \\
    &\quad - \textsf{E}_q[\log q(\phi, \pmb v, \pmb L, \pmb Z)].
\end{align*}
Maximizing this bound minimizes the KL divergence between the variational approximation and the true posterior, ensuring an optimal fit. Each term in the bound captures a specific contribution: Prior distributions for the latent variables and parameters, assignments of latent variables to observed data, the likelihood of the observed data, and the complexity of the variational posterior. This decomposition provides valuable insight into the trade-offs involved in the approximation process, guiding the optimization to a balance between computational feasibility and the accuracy of the posterior estimate.

\subsubsection{Data, priors, and performance assessment}

The dataset used in this analysis is the \texttt{galaxies} dataset from the \texttt{MASS} library \citep{VenablesRipley2002} in \texttt{R}, consisting of measurements of the velocities of 82 galaxies. These velocities are recorded in kilometers per second (km/s) and are corrected for the expansion of the universe (recessional velocities). Pre-processing involves standardizing the data to ensure a mean of zero (\(\bar{y} = 0\)) and a standard deviation of one (\(s_y = 1\)).

For prior specification, a non-informative prior is used for the location parameter, with \(\psi = 0\). The scale parameter is determined based on the decomposition of the sample variance. Specifically, the expected variance of the observations is given by:
\[
\textsf{Var}(y) = \textsf{E}[\textsf{Var}(y \mid \theta)] + \textsf{Var}[\textsf{E}(y \mid \theta)],
\]
where \(y \mid \theta \sim \textsf{N}(\theta, 1/\phi)\). These terms are given by \(\textsf{E}(1/\phi) + \textsf{Var}(\theta)\), respectively, with \(\phi \sim \textsf{G}(a_\phi, b_\phi)\) and \(\theta \sim \textsf{N}(\psi, 1/\nu)\). Substituting these distributions, the decomposition becomes:
\[
\textsf{E}[\textsf{Var}(y)] = \frac{b_\phi}{a_\phi - 1} + \nu^{-1}.
\]
To encourage smoothness in the estimates, small values of \(1/\phi\) are prioritized. An allocation ratio of \(1:7\) is used between observation-level and system-level variances. The hyperparameters are set as follows:
\[
a_\phi = 1.5, \quad b_\phi = 0.5 \cdot \frac{1}{8}, \quad \nu = \left(\frac{7}{8}\right)^{-1}.
\]

For benchmarking, a collapsed Gibbs sampler (e.g., \citealt{rodriguez2013nonparametric}) is implemented. The sampler is configured with 1,000 burn-in iterations, followed by 10,000 posterior samples to ensure reliable inference.

\subsubsection{Results}

\begin{figure}[!htb]
    \centering
    \includegraphics[scale=0.7]{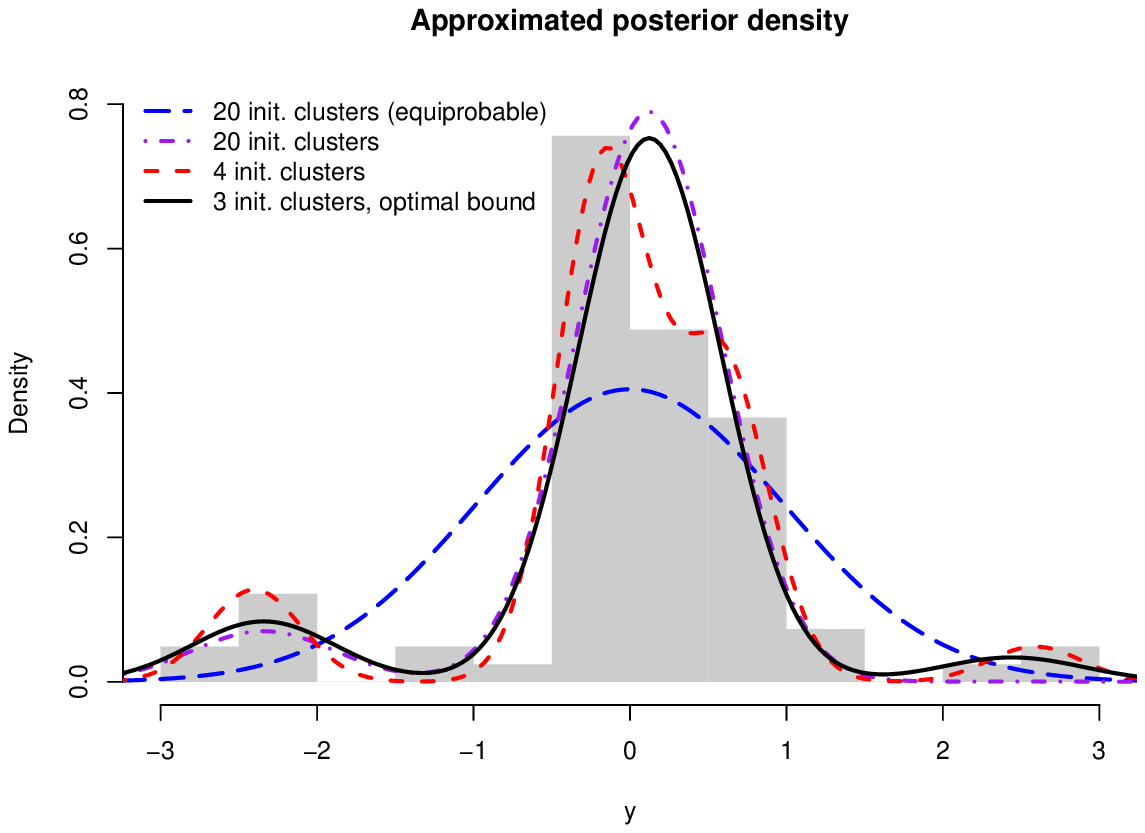}
    \caption{Variational approximations with different initial settings.}\label{fig_var_1}
\end{figure}

Different initial settings were tested to evaluate their impact on the variational algorithm's performance (Table \ref{tab_var_1} and FIgure \ref{fig_var_1}). Among the scenarios considered, assigning data into three initial clusters resulted in the optimal variational bound, indicating that this configuration provides the best initialization for the algorithm.

\begin{table}[!htb]
    \centering
    \begin{tabular}{c c c c c}
    \hline
    \multirow{2}{*}{Scenario}  &  Initial & \# of Iterations & Running & Lower Bound on \\
     &  Setting & to Convergence & Time (sec) & Log-Marginal Prob.\\
    \hline
    1 & 20 initial clusters & 101 & ~6.88 & -268.1 \\
    2 & 20 initial clusters & 216 & 14.85 & -275.6 \\
    3 & ~4 initial clusters & ~65 & ~4.81 & -274.0 \\
    \textbf{4} & ~\textbf{3 initial clusters} & ~\textbf{11} & ~\textbf{0.84} & \textbf{-245.7} \\
    \hline
    \end{tabular}
    \caption{Variational approximation with different initial settings.}\label{tab_var_1}
\end{table}

The running time comparison highlights the efficiency of the variational algorithm (VA) compared to the collapsed Gibbs sampler (CGS). The variational algorithm completed in 0.84 seconds, whereas the collapsed Gibbs sampler required 637.8 seconds, demonstrating a significant improvement in computational speed, particularly on a \textbf{\(10^3\)-scale}.

\begin{figure}[!htb]
    \centering
    \subfigure[Predictive density.]{\includegraphics[width = 0.49\textwidth]{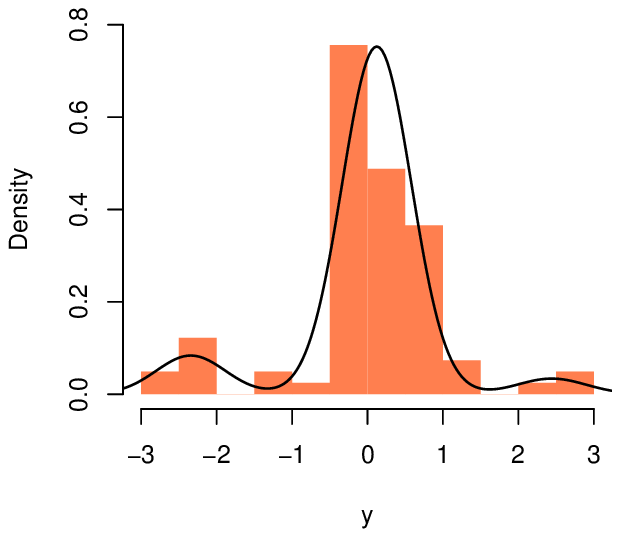}}
    \subfigure[Heatmap of incidence matrix.]{\includegraphics[width = 0.49\textwidth]{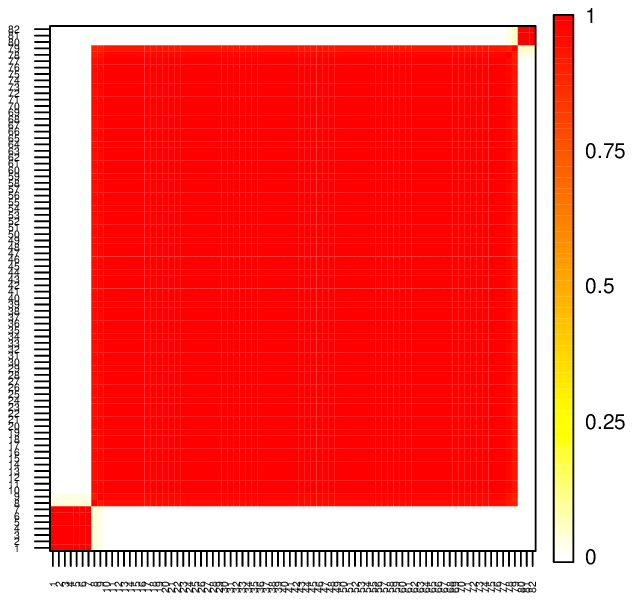}}
    \caption{Posterior samples from the variational algorithm (initial setting 4).}\label{fig_var_2}
\end{figure}

For the variational algorithm, posterior samples were obtained using initial setting 4. The results include the predictive density and the heatmap of the incidence matrix (a matrix of co-clustering probabilities indicating the probability that two data points belong to the same cluster), as shown in \ref{fig_var_2}. Subfigure (a) presents the predictive density, while Subfigure (b) displays the heatmap of the incidence matrix, revealing the clustering structure inferred by the algorithm.

\begin{figure}[!htb]
    \centering
    \subfigure[Predictive density.]{\includegraphics[width = 0.49\textwidth]{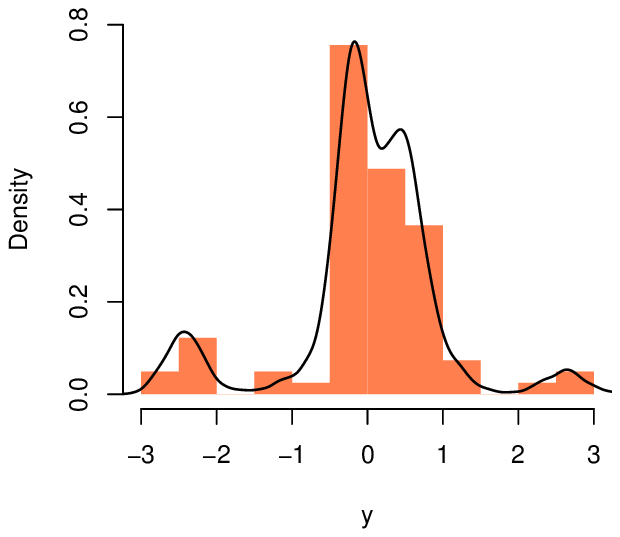}}
    \subfigure[Heatmap of incidence matrix.]{\includegraphics[width = 0.49\textwidth]{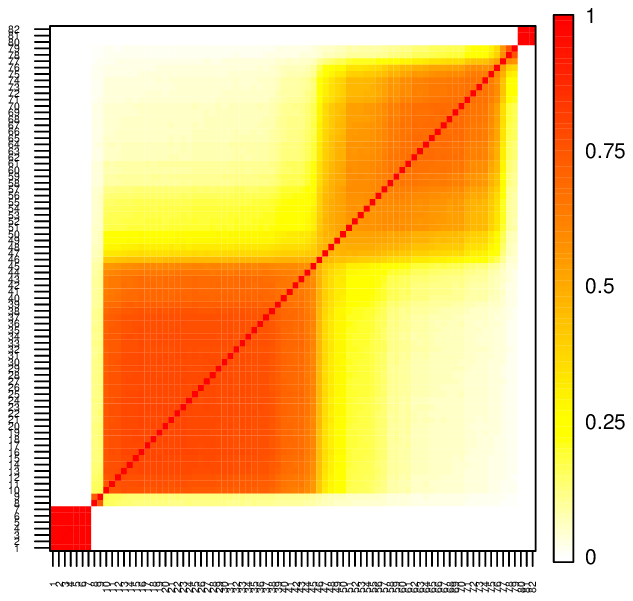}}
    \caption{Posterior samples from the collapsed Gibbs sampler.}\label{fig_var_3}
\end{figure}

For the collapsed Gibbs sampler, posterior samples were similarly obtained. Figure \ref{fig_var_3} presents the corresponding results: Subfigure (a) shows the predictive density, and Subfigure (b) displays the heatmap of the incidence matrix. These visualizations demonstrate the clustering structure identified by the Gibbs sampler.

\subsubsection{Remarks}

We compare the VA and the CGS based on their performance in fitting the data. Both methods demonstrated a reasonable fit, but their behavior differed in several key aspects. In terms of computational efficiency, the VA outperformed the CGS, with a running time difference on the order of \(10^3\), highlighting the VA's much faster performance. However, the covariance pattern between parameters, which was captured effectively by the CGS, was lost in the VA. Additionally, the CGS exhibited sensitivity to prior specification, while the VA was relatively insensitive, making it more robust to prior choices.

The VA offers several advantages that make it a compelling choice for large-scale or complex problems. It is highly computationally efficient, with truncation occurring at the approximation level (variational distribution) rather than at the system level (prior specification). This property contributes to its robustness against prior choices and its rapid and quantifiable convergence. Moreover, the performance of the VA is quantifiable using metrics such as the KL divergence, providing insights into the quality of the approximation. Its iterative approximation approach is easier to implement compared to sampling-based methods like MCMC.

However, the VA has notable drawbacks. It loses covariance information between parameters, which may be critical in certain applications. It is highly sensitive to initial settings, with a strong tendency toward local convergence, which can affect the quality of the inference. Additionally, its inference accuracy is generally lower than that of sampling-based methods. Furthermore, the methodology for handling parameters outside the exponential family is unclear, limiting its generalizability in some contexts. The most challenging aspect of implementing the VA is specifying the initial settings. Since the algorithm's performance and convergence depend heavily on the initialization, careful consideration is required to achieve reliable results.

\section{Discussion}

The results presented in this paper highlight the broad utility of Bayesian inference across diverse domains, ranging from hierarchical and spatial modeling to non-parametric approaches such as Dirichlet process mixtures and mixtures of Brownian motion. By leveraging both Markov chain Monte Carlo (MCMC) methods and variational approximations, this work demonstrates the dual strengths of Bayesian inference: its capacity for accurate, detailed posterior analysis and its adaptability to large-scale problems requiring efficient computation.

MCMC methods remain the gold standard for Bayesian computation due to their ability to accurately sample from complex posterior distributions. Applications in spatial modeling and higher-order Markov chains particularly benefit from the flexibility of MCMC to incorporate intricate dependencies and hierarchical structures. However, their high computational cost and sensitivity to tuning parameters pose challenges, especially for high-dimensional or large-scale datasets. On the other hand, variational approximations provide a complementary approach, offering rapid convergence and scalability by reformulating posterior inference as an optimization problem. Although this efficiency comes at the cost of some approximation accuracy, it makes variational methods indispensable for applications where computational constraints are paramount.

The comparative analysis of MCMC and variational methods underscores their trade-offs. While MCMC excels in capturing complex covariance structures and delivering high-fidelity posterior estimates, it often requires significant computational resources and careful tuning. In contrast, variational methods are less sensitive to priors and achieve rapid convergence, but they may oversimplify posterior distributions by neglecting parameter correlations. These differences highlight the importance of method selection based on the specific requirements of the problem, including the dataset size, model complexity, and computational resources available.

The applications explored in this paper emphasize the versatility of Bayesian methods across scientific fields. For example, spatial modeling in climatology and epidemiology underscores the power of Bayesian inference to capture geographic dependencies and uncertainty, while hierarchical models in oceanography and finance demonstrate its ability to handle nested and multi-level data structures. Non-parametric methods such as Dirichlet process mixtures reveal Bayesian inference’s flexibility in modeling infinite-dimensional parameter spaces, making it suitable for adaptive clustering and density estimation.

Despite their strengths, Bayesian methods face ongoing challenges. The sensitivity of MCMC methods to initialization and tuning parameters can limit their accessibility to non-expert users. Similarly, the reliance of variational approximations on strong assumptions about the posterior distribution can lead to biased inference in cases involving highly complex models. Future advancements should focus on developing hybrid methods that combine the strengths of MCMC and variational inference, such as stochastic variational techniques, to improve both accuracy and scalability.

Looking ahead, Bayesian inference is poised to benefit from integration with modern computational tools, including parallel processing, machine learning, and GPU-based computation, which promise to enhance scalability and reduce runtime. Additionally, developing methodologies to extend Bayesian inference to more complex, non-standard problems will further expand its applicability. This work aims to bridge the gap between theory and practice, providing a foundation for researchers to explore Bayesian methods across a diverse range of applications and challenges.

\section*{Statements and Declarations}

The authors declare that they have no known competing financial interests or personal relationships that could have appeared to influence the work reported in this article.

During the preparation of this work the authors used ChatGPT-4-turbo in order to improve language and readability. After using this tool, the authors reviewed and edited the content as needed and take full responsibility for the content of the publication.

\bibliography{references.bib}
\bibliographystyle{apalike}

\appendix

\section{Notation}

The cardinality of a set \(A\) is denoted by \(|A|\). If \(P\) is a logical proposition, then \(I(P) = 1\) if \(P\) is true and \(1_{\text{P}} = 0\) if \(P\) is false. The Gamma function is defined as \(\Gamma(x) = \int_0^\infty u^{x-1} e^{-u} \, \text{d}u\). Matrices and vectors whose entries consist of subscripted variables are represented using bold notation. For example, \(\pmb{x} = (x_1, \dots, x_n)\) denotes an \(n \times 1\) column vector with elements \(x_1, \dots, x_n\). We use \(\pmb{0}\) and \(\boldsymbol{1}\) to denote column vectors with all entries equal to 0 and 1, respectively, and \(\mathbf{I}\) to represent the identity matrix. A subscript in this context indicates the corresponding dimension; for instance, \(\mathbf{I}_n\) refers to the \(n \times n\) identity matrix. The transpose of a vector \(\pmb{x}\) is denoted by \(\pmb{x}^\top\), and the notation extends analogously to matrices. Additionally, for a square matrix \(\mathbf{X}\), we use \(\text{tr}(\mathbf{X})\) to denote its trace and \(\mathbf{X}^{-1}\) for its inverse. The norm of \(\pmb{x}\), given by \(\sqrt{\pmb{x}^\top \pmb{x}}\), is denoted by \(\|\pmb{x}\|\).

\end{document}